\documentclass[graybox,vecphys]{svmult}

\usepackage{makeidx}         
\usepackage{graphicx}        
\usepackage{epsfig}

\usepackage{multicol}        
\usepackage[bottom]{footmisc}
\usepackage{xcolor}

\usepackage{newtxtext}       %
\usepackage{newtxmath}       

\usepackage{url} 

\makeindex             

\begin{document}

\newcommand{\jcmindex}[2]{\index{{\bf\large #1}!#2}}
\newcommand{\jcmindext}[3]{\index{{\bf\large #1}!#2!#3}}

\title*{Nonlinearity and Topology}
\author{
Avadh Saxena
\and
Panayotis G.\ Kevrekidis
\and
Jes\'us Cuevas-Maraver
}

\institute{
Avadh Saxena
\at
Theoretical Division and Center for Nonlinear Studies,\\
Los Alamos National Laboratory,\\
Los Alamos, New Mexico 87545, USA\\
\email{avadh@lanl.gov}
\and
Panayotis G. Kevrekidis
\at
Department of Mathematics and Statistics,\\
University of Massachusetts,\\
Amherst, MA 01003-4515, USA\\
\email{kevrekid@umass.edu}
\and
Jes\'us Cuevas-Maraver
\at
Grupo de Fisica No Lineal,\\
Departamento de Fisica Aplicada I, Universidad de Sevilla,\\
Escuela Polit\'ecnica Superior, C/ Virgen de \'Africa, 7\\
41011 Sevilla, Spain
\and
Jes\'us Cuevas-Maraver
\at
Instituto de Matem\'aticas de la Universidad de Sevilla (IMUS)\\
Edificio Celestino Mutis,\\
Avda. Reina Mercedes s/n\\
41012 Sevilla, Spain\\
\email{jcuevas@us.es}
}
\maketitle
\abstract
{
The interplay of nonlinearity and topology results in many novel and emergent properties across a number of physical systems such as chiral magnets, nematic
liquid crystals, Bose-Einstein condensates, photonics, high energy physics, etc.  It also results in a wide variety of topological defects such as solitons, vortices,
skyrmions, merons, hopfions, monopoles to name just a few. Interaction among and collision of these nontrivial defects itself is a topic of great interest.  Curvature
and underlying geometry also affect the shape, interaction and
behavior of these
defects. Such properties can be studied using techniques such as,
e.g. the Bogomolnyi decomposition.
Some applications of this interplay, e.g. in nonreciprocal photonics as well as topological materials such as Dirac and Weyl semimetals, are also elucidated. }


\section{Introduction}

The main context of this chapter is how topological effects in nonlinear systems give rise to a rich playground of excitations and properties.
Topology, whether in real space or momentum space or more generally in a parameter space, is associated with certain system properties
remaining unaltered under continuous deformation.  It follows that during deformation neighboring points remain close to each other.
Topology could be local, e.g. change in the lattice or network due to a defect, or global.  The latter means attributes such as the genus ($g$)
or Euler characteristic ($\chi$) are overall or global features of a system.

Boundary conditions play an important role (through $\chi$), for example in finite carbon nanotubes or {\it edge states} in topological materials: quantum
Hall systems, topological insulators \cite{TI}, topological superconductors \cite{topsup}, Dirac and Weyl semimetals \cite{semimetals}, etc. The latter are
three dimensional analogs of graphene featuring gapless electronic excitations that are protected by topology and (time reversal, space inversion or
other crystalline) symmetry \cite{semimetals}. Very recent experiments indicate that these materials, specifically Weyl semimetals, may also provide a
realization of axions, very weakly interacting neutral particles in quantum field theory and potential candidates for dark matter, in condensed matter
\cite{axion1, axion2}.

In many physical systems and materials \cite{topobook} there are point defects as well as extended or topological defects.  The topological defects can significantly
alter the physical properties and dynamics of the system.  Apart from the celebrated soliton-like defects there is a whole slew of more elaborate ones that
include skyrmions, merons, hopfions, monopoles, dislocations, disclinations among others.  In this chapter we discuss how such defects arise in chiral
magnets, nematic liquid crystals, Bose-Einstein condensates (BECs), etc.  We also discuss the role of topology in the momentum space, particularly in the
context of topological materials.

The combination of nonlinearity and topology also provides a highly desirable functionality in photonics, namely nonreciprocity, which is quite important for a
variety of photonic devices including optical isolators \cite{nonreciprocity}. We thus provide examples of the interplay between nonlinearity and topology in
photonics as well as condensed matter analogs. In addition, we
illustrate the role of geometry and topology in determining spin
textures via the so-called Bogomolnyi
decomposition \cite{bogom}.  Finally, we discuss several open problems and future directions with regard to the role of topology in the presence of nonlinearity.


\section{Topological defects in Nonlinear field theories}
\label{sec:topdef}

We consider a variety of topological defects that arise in a number of field theories including the nonlinear $\sigma$-model \cite{MantonSut, ShnirBook}.  The
Hamiltonian for the latter is given by
\begin{equation}\label{eq:sigma}
H = \int (\nabla{\mathbf n})^2 d^2 x \,, ~~~ {\mathbf n}^2 = 1 \,,
\end{equation}
where the unit vector ${\mathbf n}$ lives on a unit sphere.  This model can support scalar soliton configurations under special conditions. However, under a
scaling transformation $x\rightarrow\lambda x$ and $y\rightarrow\lambda y$ the Hamiltonian $H$ remains invariant and thus a soliton solution can be
trivially scaled to a point because there is no length scale in the plane.  We will return to this point later when we consider the nonlinear $\sigma$-model on
curved manifolds.

We note here that to describe disordered Weyl semimetals an anisotropic topological term can be analytically derived from the action of the nonlinear
$\sigma$ model \cite{disorder}. In the next section we consider a
variety of topological defects such as skyrmions, merons and hopfions.
Given their extensive interest and applicability, subsequently in
Sec. 4, we consider vortices and vortex loops/rings. We then turn
to different prototypical applications such as liquid crystals
(and the emergence of skyrmions in them) in Sec. 5, as well
as Bose-Einstein condensates in Sec. 6. After providing
an example of a theoretical tool for the study of topology in curved
manifolds via the Bogomolnyi decomposition (section 7), we present a broader
perspective of the impact of topological ideas in Materials (Sec. 8),
Optics (Sec. 9) and Acoustics and beyond (Sec. 10).
Then in Sec. 11, we summarize our findings and present
our Conclusions, as well as some directions for future work.

\section{Skyrmions, Merons and Hopfions}\jcmindex{S}{Skyrmion}

Beyond the well known solitons, there are more exotic topological defects such as vector field or spin textures
called skyrmions \cite{ShnirBook}, which can have topological charge
of one (or two or even more).  As shown in Fig. \ref{fig1}, in a
skyrmion at the outer boundary all spins point up (red arrows) whereas at the center there is a spin pointing
down (blue arrow).  Therefore somewhere in the middle the spins have
to lie in the plane (green arrows). Half skyrmions are also referred
to as
merons and have a topological
charge of one half: the outer spins point up whereas the spin in the center lies in the plane.  There are many other related topological defects such as sphalerons and bags
(or lumps) known in high energy physics \cite{MantonSut}.  Similarly,
three dimensional defects, e.g. vortex lines, vortex loops (rings)
and knots appear in many physical systems, including,
e.g. polymeric knots \cite{Muthukumar}.

\subsection{Skyrmions in Chiral Magnets}
\label{sec:sky} \jcmindex{C}{Chiral magnets}
Before discussing magnets we note that beyond magnetic materials (e.g. ferroelectrics) skyrmions have been observed at interfaces \cite{das}. In this case, the texture is given in terms of polar vectors or electric dipoles.  Nanodots and nanocomposites can also stabilize skyrmions (due to boundary conditions) in polar materials \cite{bellaiche}.

The Hamiltonian for a chiral magnet (i.e. lacking spatial inversion symmetry in its crystal structure, e.g. MnSi) consists of the nonlinear $\sigma$-model plus the Zeeman
term in addition to the Dzyaloshinskii-Moriya interaction \cite{nagaosa}
\begin{equation}\label{eq2}
H = \int \left[J(\nabla{\mathbf n})^2 + D{\mathbf n}\cdot(\nabla\times{\mathbf n}) - {\mathbf n}\cdot {\mathbf B} \right]d^2 x \,,
\end{equation}
where $J$ denotes the (magnetic) exchange constant, ${\mathbf B}$ is the external magnetic field in the last term in the Hamiltonian representing the Zeeman interaction
and $D$ represents the strength of the Dzyaloshinskii-Moriya interaction.  The latter arises from the spin-orbit interaction at the microscopic level.  Here ${\bf n}$
is a unit vector describing the direction of the magnetic moment.

The topological charge associated with a skyrmion spin configuration is given by \cite{nagaosa} \jcmindex{T}{Topological charge}
\begin{equation}
Q=\frac{1}{4\pi}\int dr^2 \left[\mathbf{n}\cdot(\partial_x\mathbf{n}\times\partial_y\mathbf{n})\right] =\pm1 \,.
\end{equation}
For a metallic material when a conduction electron traverses across a skyrmion it gets spin polarized as a result of its interaction with the spin configuration of the
skyrmion.  In addition, the conduction electron is subjected to an effective electric and magnetic field \cite{nagaosa}, respectively given by
\begin{eqnarray}\label{eq3}
E&=&\frac{\hbar}{2e}[\mathbf{n}\cdot(\nabla\mathbf{n}\times\partial_t\mathbf{n})] , \label{eq3a}\\
B&=&\frac{\hbar c}{2e}[\mathbf{n}\cdot(\partial_x\mathbf{n}\times\partial_y\mathbf{n})] \label{eq3b}.
\end{eqnarray}
These emergent fields give rise to the topological and skyrmion Hall effects.

\begin{figure}[t]
\centering
\includegraphics[width=1.0\textwidth]{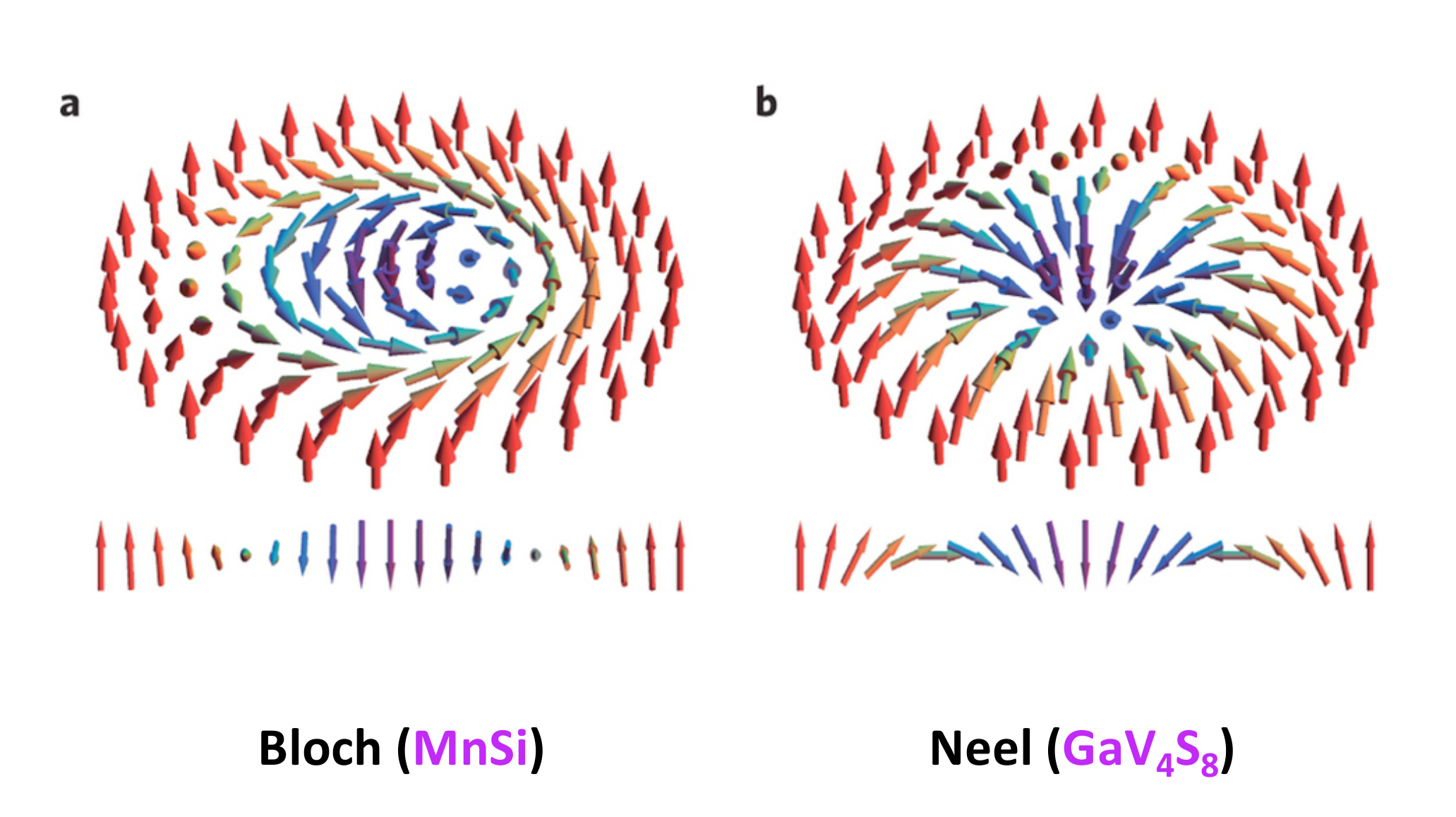}
\caption{(a) Bloch and (b) N\'eel type skyrmions in a chiral magnet MnSi and a lacunar spinel  GaV$_4$S$_8$, respectively. Below the skyrmion textures their radial cross sections show Bloch and N\'eel wall, respectively. Reproduced from \cite{NeelSkyrmion}. \copyright 2015 by the Nature Publishing Group}
\label{fig1}
\end{figure}

\begin{figure}[t]
\centering
\includegraphics[width=1.0\textwidth]{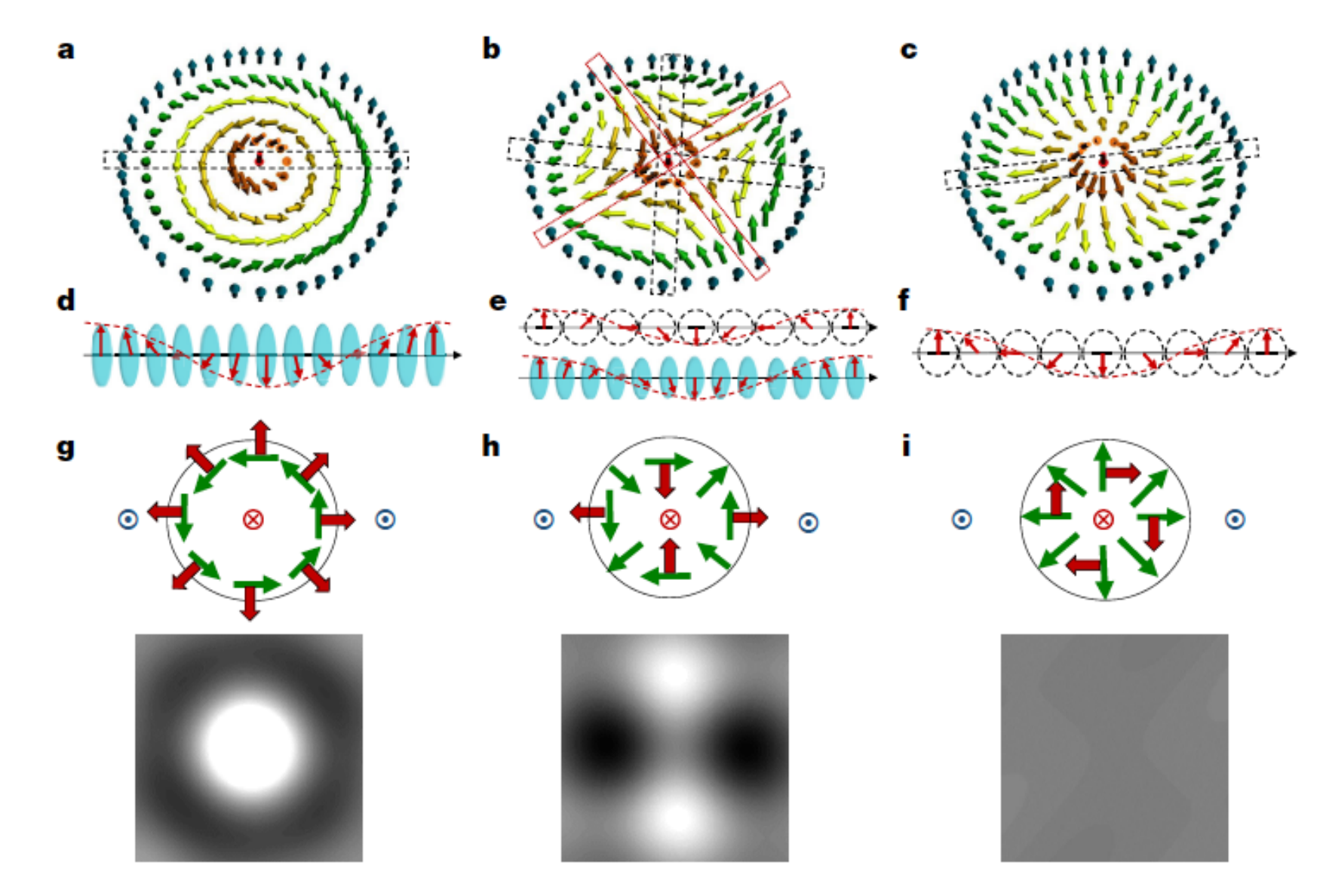}
\caption{(a) Anti-skyrmions (center panels) in
  Mn$_{1.4}$Pt$_{0.9}$Pd$_{0.1}$Sn and its comparison with the Bloch
  (left panels) and N\'eel skyrmions (right panels). The radial cross
  sections are shown below the spin textures.  Schematics of the
  magnetic moments (green arrows) and the associated Lorentz
  deflections of transmitted electrons (red arrows) are shown in the
  middle row.  The bottom row shows the corresponding
  simulated Lorentz Transmission Electron Microscopy (LTEM) patterns with dark and bright lobes. Reproduced from \cite{antiskyrmion}. \copyright 2017 by the Nature Publishing Group}
\label{fig2}
\end{figure}

Depending on the material or system, skyrmions can have chirality (i.e. Bloch skyrmion as in MnSi) or no chirality (i.e. N\'eel skyrmion as in GaV$_4$S$_8$) as depicted in Fig. \ref{fig1}.
They are so called because the radial cut of a Bloch skyrmion provides a magnetic Bloch domain wall (which is a strictly 3D structure) whereas a radial cut of the N\'eel skyrmion
leads to a magnetic N\'eel domain wall (which is a 2D structure).
Note that the spin configurations of the two types of skyrmions are
topologically equivalent.
Anti-skyrmions, which
have structural characteristics of both the Bloch and N\'eel skyrmions, have also been observed using Lorentz Transmission Electron Microscopy (LTEM) \cite{antiskyrmion} in
tetragonal Heusler materials even above the room temperature, as shown in Fig. \ref{fig2}. We note here that spin-1 photonic skyrmions have also been described in the literature
\cite{photonic}.

\subsection{Merons}
\label{sec:mer} \jcmindex{M}{Meron}
Half-skyrmions whose field covers only a hemisphere (i.e. topological charge 1/2) are known as merons. They have been observed (and modeled) in liquid crystals
\cite{liquidcrystal} and magnetic multi-layers \cite{wintz}. If we add a magnetic anisotropy energy term $A n_z^2$ to the skyrmion Hamiltonian in Eq. (2), in the large anisotropy limit a skyrmion breaks into merons \cite{merons}.  When $A>0$ it is called the easy-plane anisotropy which is what we will consider here.   The case of $A<0$ is called the easy-axis anisotropy.  As $A$ is increased the skyrmion size increases, particularly due to the expansion of the equatorial region of the skyrmion.  At a certain large value of $A$, skyrmions become unstable and merons emerge.

\begin{figure}[t]
\centering
\includegraphics[width=1.0\textwidth]{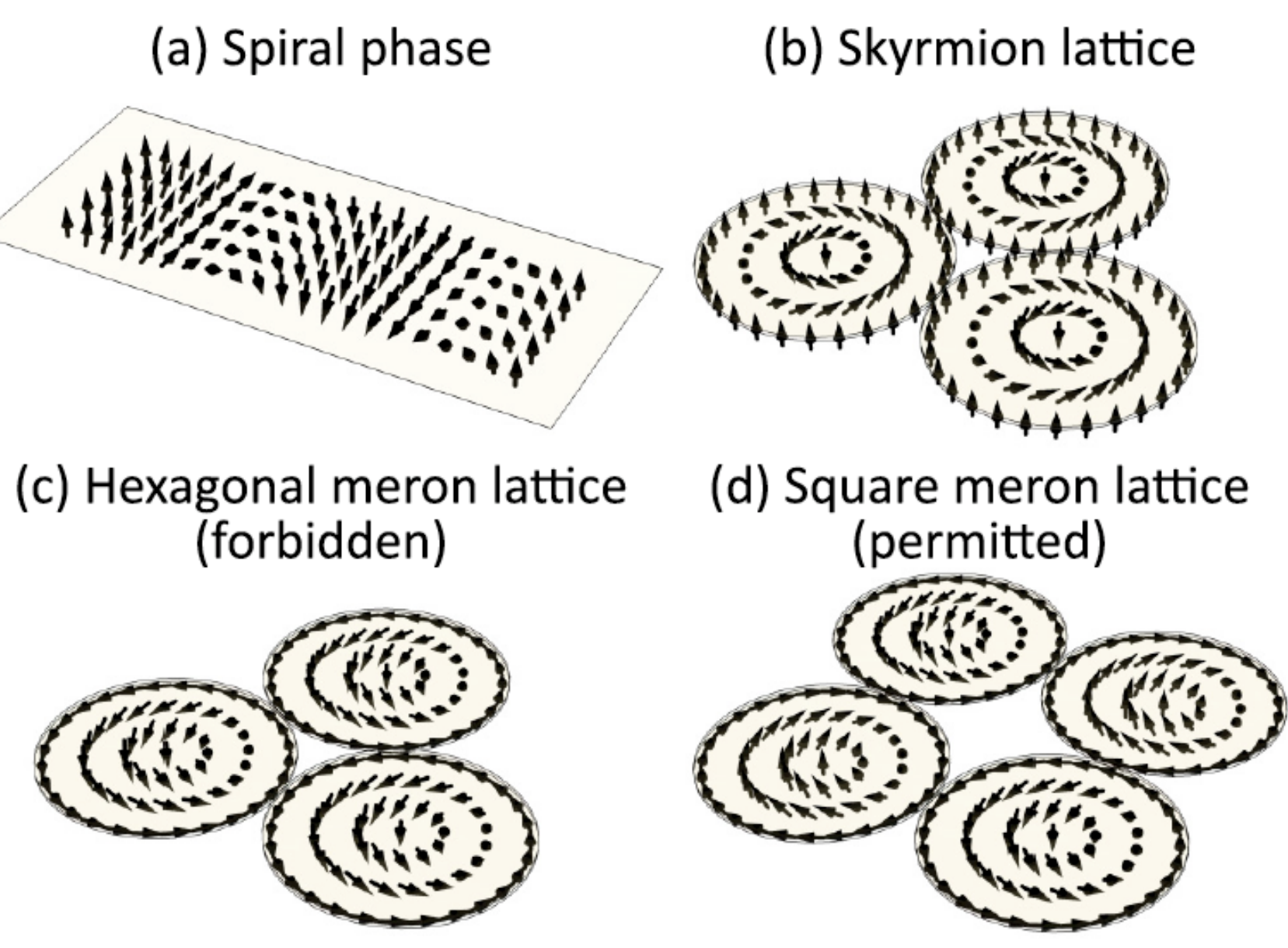}
\caption{Comparison of skyrmion and meron latices in chiral magnets. A triangular skyrmion lattice arises from a spiral phase (top row) but a triangular meron lattice is at best metastable (bottom left).  In contrast, a square meron lattice is stable (bottom right) but a square skyrmion lattice is at best metastable. On the other hand, in nematic liquid crystals a triangular meron lattice is stable but a triangular skyrmion lattice is unstable (not shown).  Reproduced from \cite{liquidcrystal}}
\label{fig3}
\end{figure}

In chiral magnets a stable triangular lattice of skyrmions emerges
from the spiral (or helical) phase as depicted in Fig. \ref{fig3}.
In the spiral phase all spins are parallel and point in the same direction at one end.  After a helical twist of $2\pi$ they come back to the original parallel state, see Fig. 3(a).
The square skyrmion lattice is at best metastable;
however, a square meron lattice is allowed.  On the contrary, in nematic liquid crystals the triangular skyrmion lattice is at best metastable but a triangular meron lattice is stable
\cite{liquidcrystal}.

\begin{figure}[t]
\centering
\includegraphics[width=1.0\textwidth]{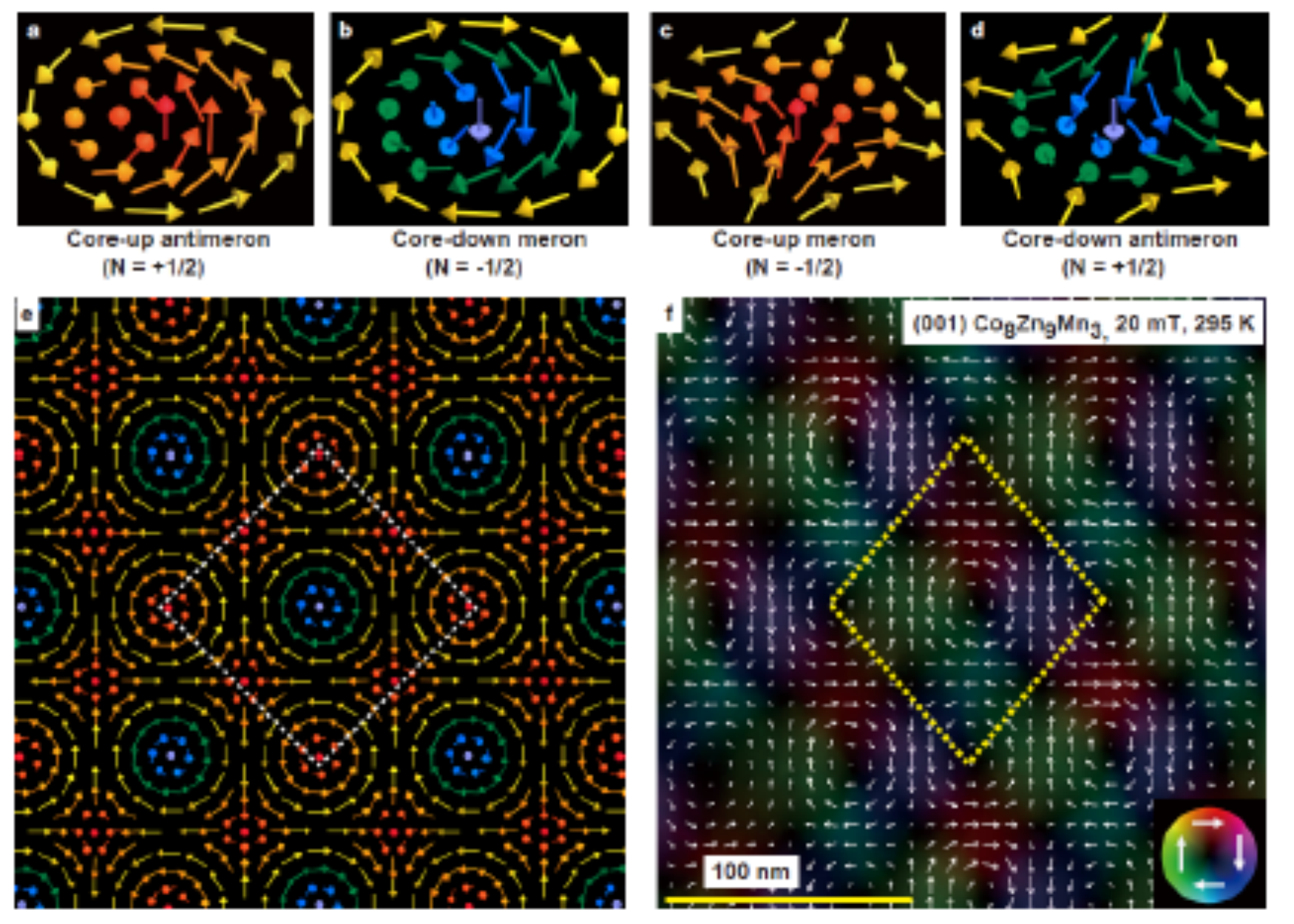}
\caption{(a) - (d) Various experimentally observed meron and anti-meron spin structures in the chiral magnet Co$_8$Zn$_9$Mn$_3$ at room temperature.  (e) Theoretically predicted and (f) LTEM observed meron/anti-meron square lattice.  Reproduced from \cite{meronsexp}}
\label{fig4}
\end{figure}

\begin{figure}[t]
\centering
\includegraphics[width=1.0\textwidth]{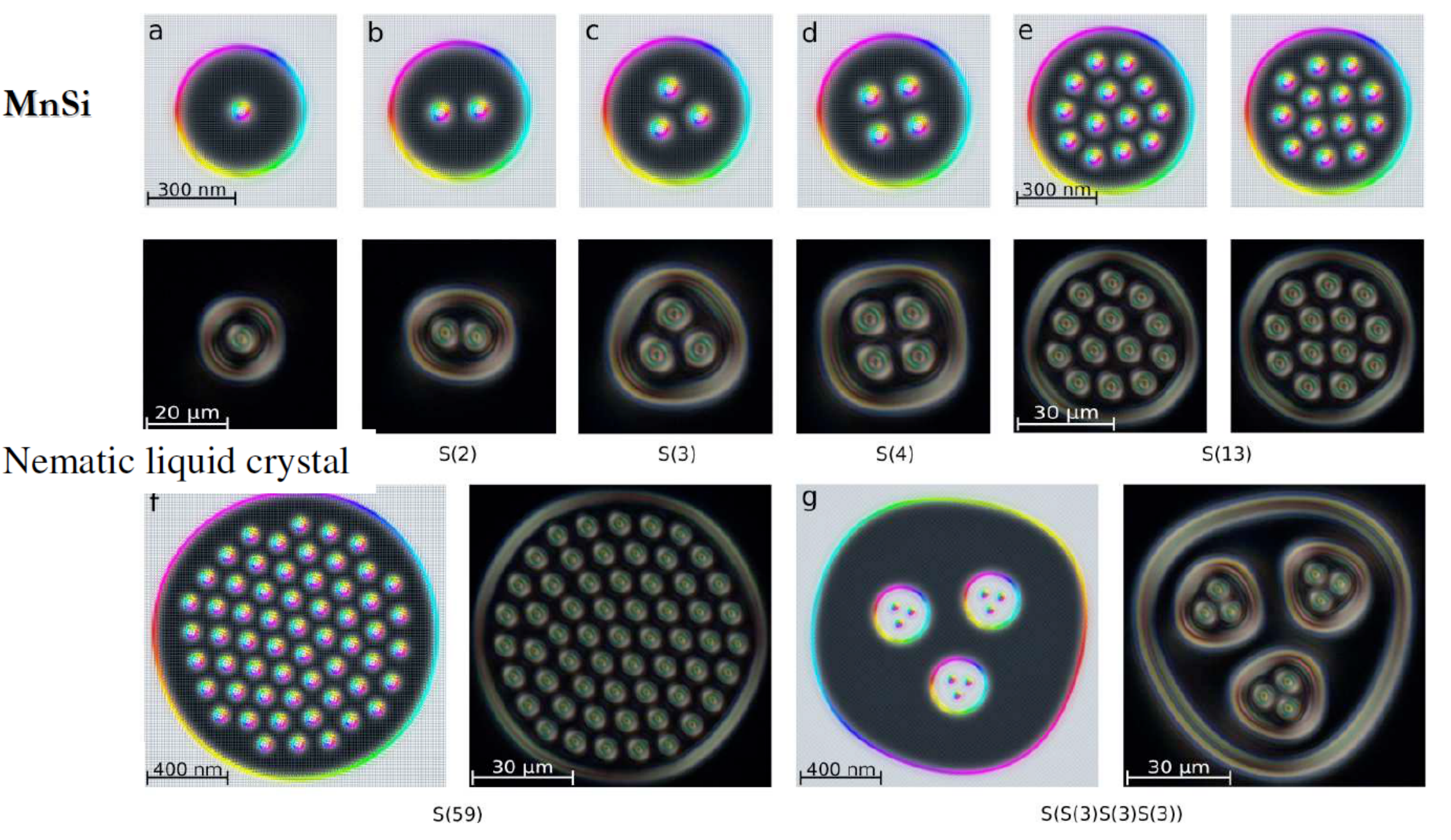}
\caption{Skyrmion bags in chiral magnets (MnSi) and nematic liquid crystals. Top panels corresponds to experimental (optical micrographs) observations in MnSi whereas middle panels are simulations of the above states. Second and fourth panels in the bottom panel are the result of simulations for the state
of the optical micrograph at their left. This figure has been adapted from the preprint version of \cite{bags} (arXiv:1806.02576v1). Only some of its panels were finally published in \cite{bags}. \copyright 2019 by the Nature Publishing Group}
\label{fig5}
\end{figure}

Merons have been experimentally observed at room temperature in a chiral-lattice magnet Co$_8$Zn$_9$Mn$_3$, a material exhibiting in-plane magnetic anisotropy \cite{meronsexp}.
In this material a meron-antimeron square lattice emerges from the helical state of spins and then transforms into a triangular lattice of skyrmions when a magnetic field is applied,
see Fig. \ref{fig4}.  Interestingly, in analogy with the baryon model
in high energy physics, skyrmion bags have been observed both in chiral magnets and nematic liquid crystals \cite{bags}.
These bags are multi-skyrmion configurations where a large skyrmion contains a variable number of antiskyrmions inside it, as depicted in Fig.~\ref{fig5}.
Merons can be compared and contrasted with magnetic vortices; there is a difference in their spin configurations.

\subsection{Hopfions and Torons}
\label{sec:hop} \jcmindex{H}{Hopfion} \jcmindex{T}{Toron}
Three dimensional topological solitons (appearing in 3+1 dimensional
scalar field theories) that can be characterized by the integer-valued
Hopf invariant are
known as hopfions \cite{ShnirBook}.  Ludwig Faddeev proposed their existence in the 1970s.  They represent one of the best known examples of knot solitons in field theory.
In 1931 Heinz Hopf considered a link of two loops, thus paving the way for the linking number of circles as a topological invariant, i.e. the Hopf number.

\begin{figure}[t]
\centering
\includegraphics[width=1.0\textwidth]{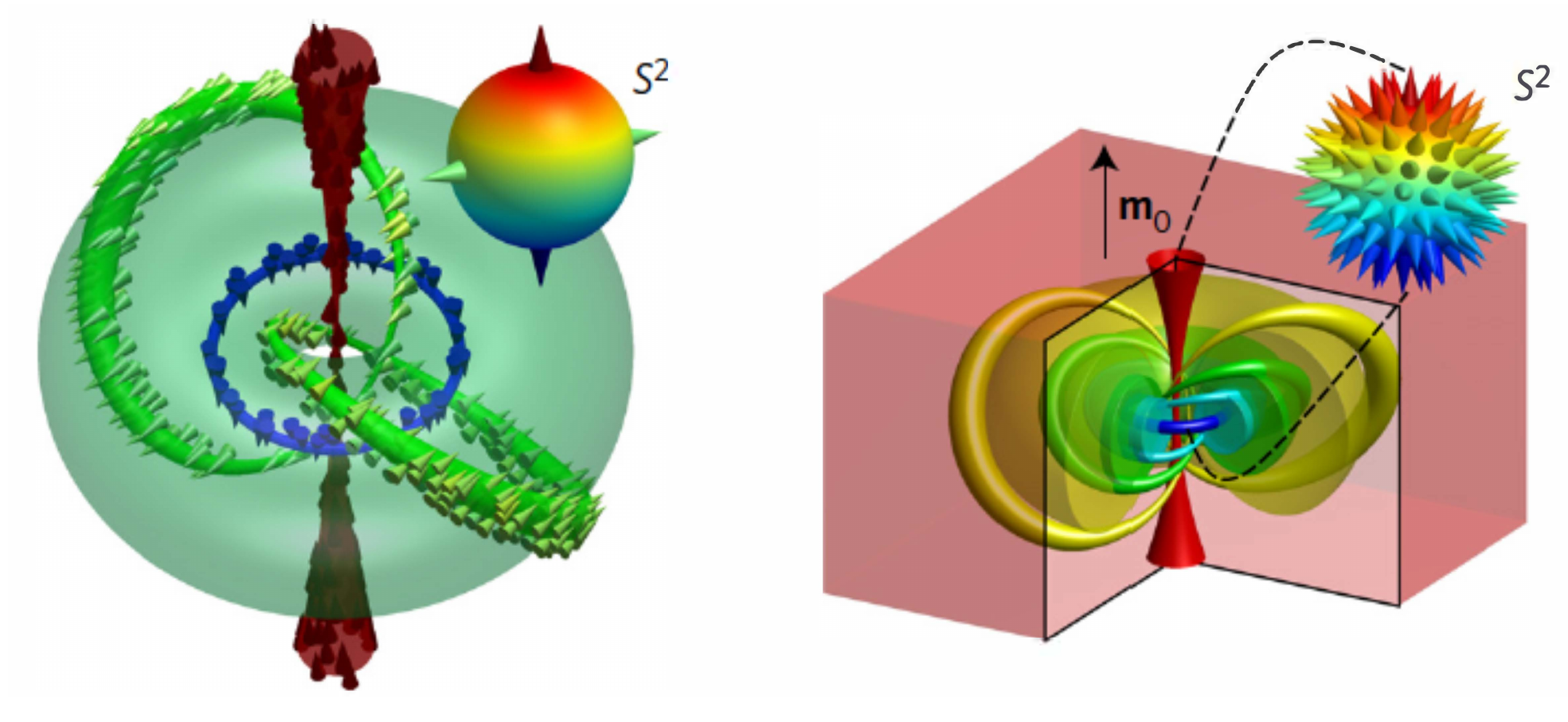}
\caption{Experimentally deduced hopfion texture in a liquid crystal (left panel) and a schematic of the hopfion as a knotted soliton (right panel).  The hopfion is obtained by linking the circle-like preimages residing on nested tori in the  material's 3D space.  The preimages correspond to color-coded points on $S^2$. Reproduced from \cite{smalyukh}. \copyright 2017 by the Nature Publishing Group}
\label{fig6}
\end{figure}

\begin{figure}[t]
\centering
\includegraphics[width=1.0\textwidth]{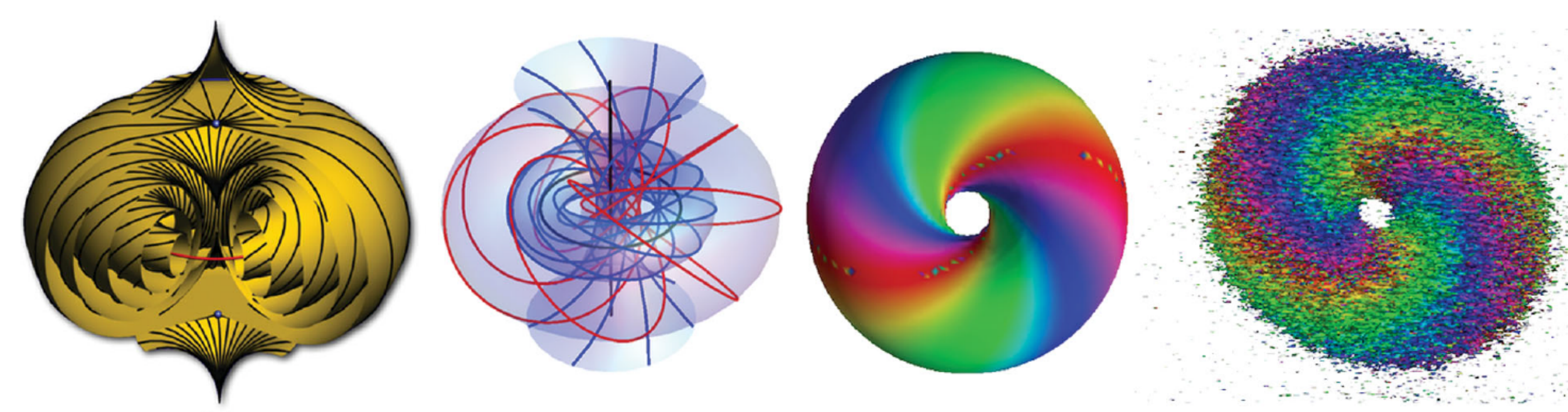}
\caption{Left panel: The texture of a toron (reproduced from \cite{toron_nmat}). \copyright 2010 by the Nature Publishing Group).  Second panel: Flow lines of the Hopf fibration.  Third panel: The preimage surface of the Hopf fibration.  Right panel: The experimental preimage of a Hopf fibration. Reproduced from \cite{fibration}. \copyright 2010 by the American Physical Society}
\label{fig7}
\end{figure}

The topological charge or linking number of a hopfion is the homotopy
group of the Hopf map $\pi_3(S^2) = {\mathbf Z}$, where ${\mathbf Z}$ is the group of relative integers. The Hopf fibration is a topologically stable
texture of a smooth, global configuration of a field.  In effect, it is an interwoven structure of preimages.  A preimage is defined as the set of all points where a field
orientation takes a specific value. The Hopf fibration has been observed in liquid crystals \cite{fibration} and so are hopfions \cite{smalyukh}, see Fig. \ref{fig6} and Fig. \ref{fig7}.
Similarly, there are light controlled torons in liquid crystals \cite{toron_nmat, torons}.  The toron is essentially a tube of double twist which is wrapped upon itself such
that its boundary forms a torus. It contains two point defects, which can be manipulated to create a defect free structure topologically equivalent to a Hopf fibration.
Finally, hopfions have also been considered in chiral magnets \cite{sutcliffe} and torus knots have been described as hopfions \cite{knots}.

\subsection{Monopoles}
\begin{figure}[t]
\sidecaption[t]
\includegraphics[width=7.5cm]{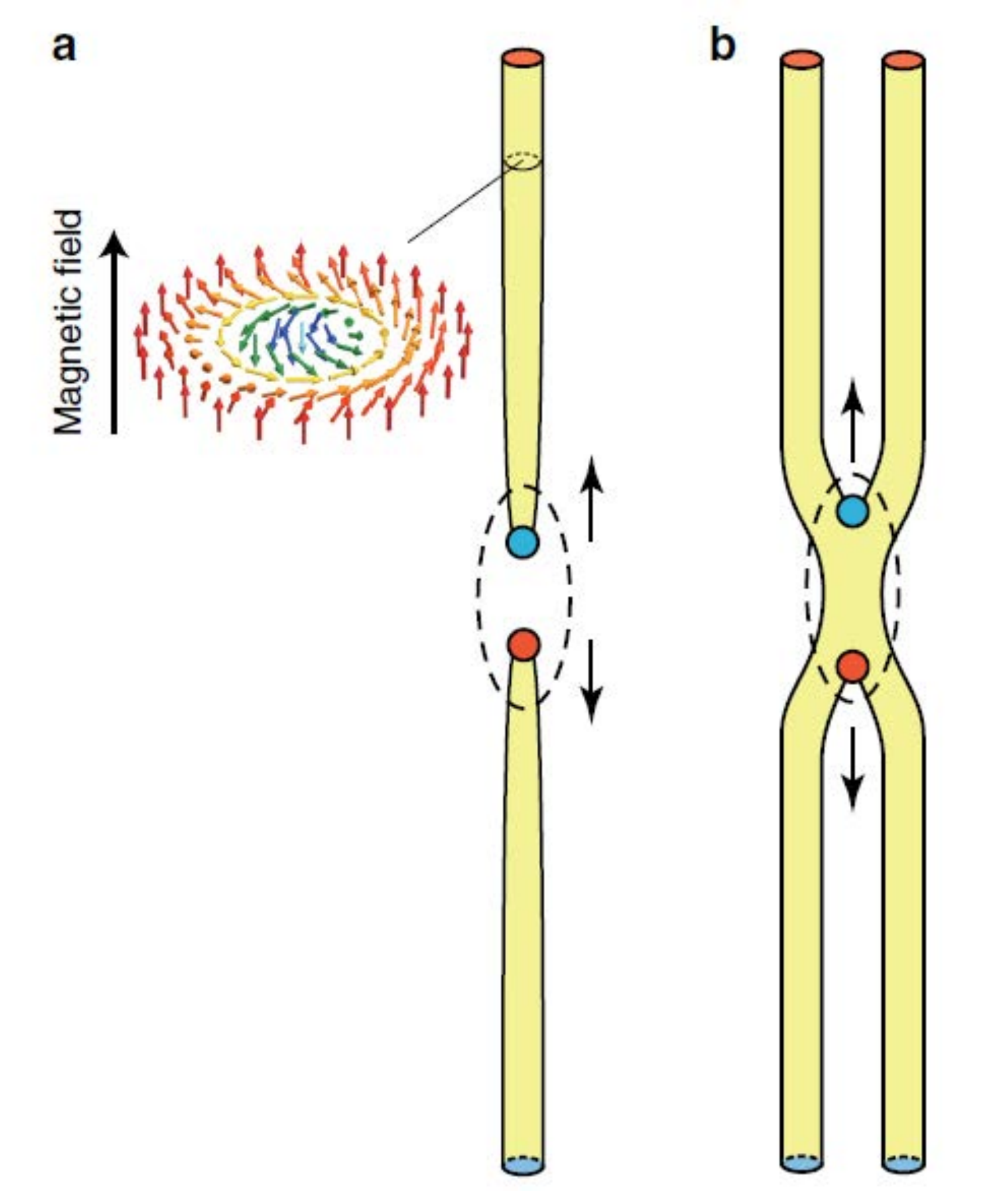}
\caption{Emergence of magnetic monopoles and anti-monopoles (colored circles).  They result either (a)  from pinching off of a skyrmion string or (b) a partial merging of two neighboring
skyrmion strings.
Reproduced from \cite{tube2}. Creative Commons Attribution License (CC BY) \protect\url{https://creativecommons.org/licenses/by/4.0/}}
\label{fig9}
\end{figure}

\begin{figure}[t]
\centering
\includegraphics[width=1.0\textwidth]{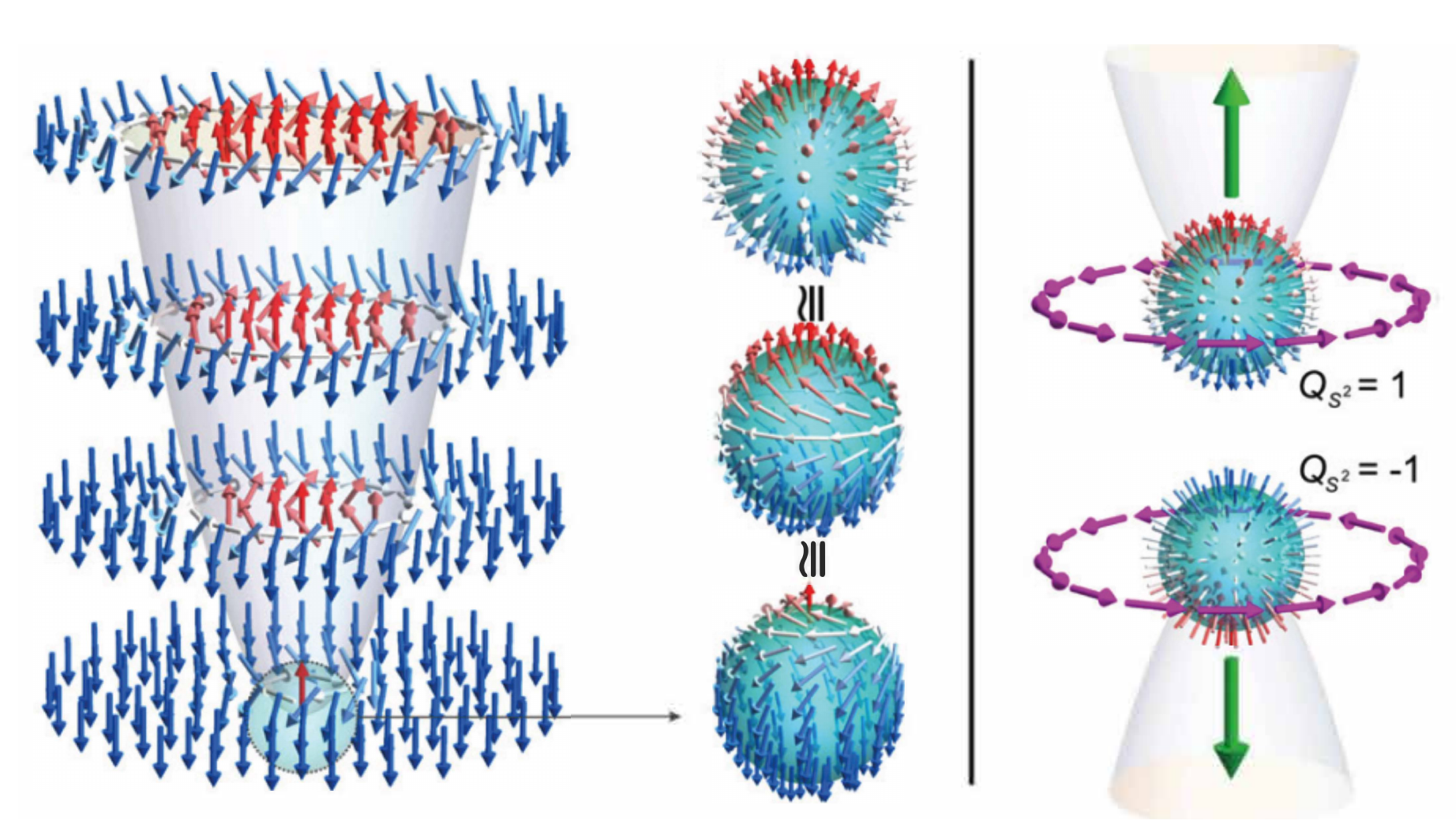}
\caption{Left panel: Spin texture at the end of a Skyrmion line. Middle panel:  The magnetization configuration is topologically equivalent to a hedgehog with radially outward magnetization, i.e. a magnetic monopole. Right panel: Two separating hedgehogs (or monopoles) with opposite topological charge with an emergent solenoidal electric field (purple arrows). Reproduced from \cite{monopole}. \copyright 2018 by the American Physical Society}
\label{fig10}
\end{figure}

\begin{figure}[t]
\centering
\includegraphics[width=0.8\textwidth]{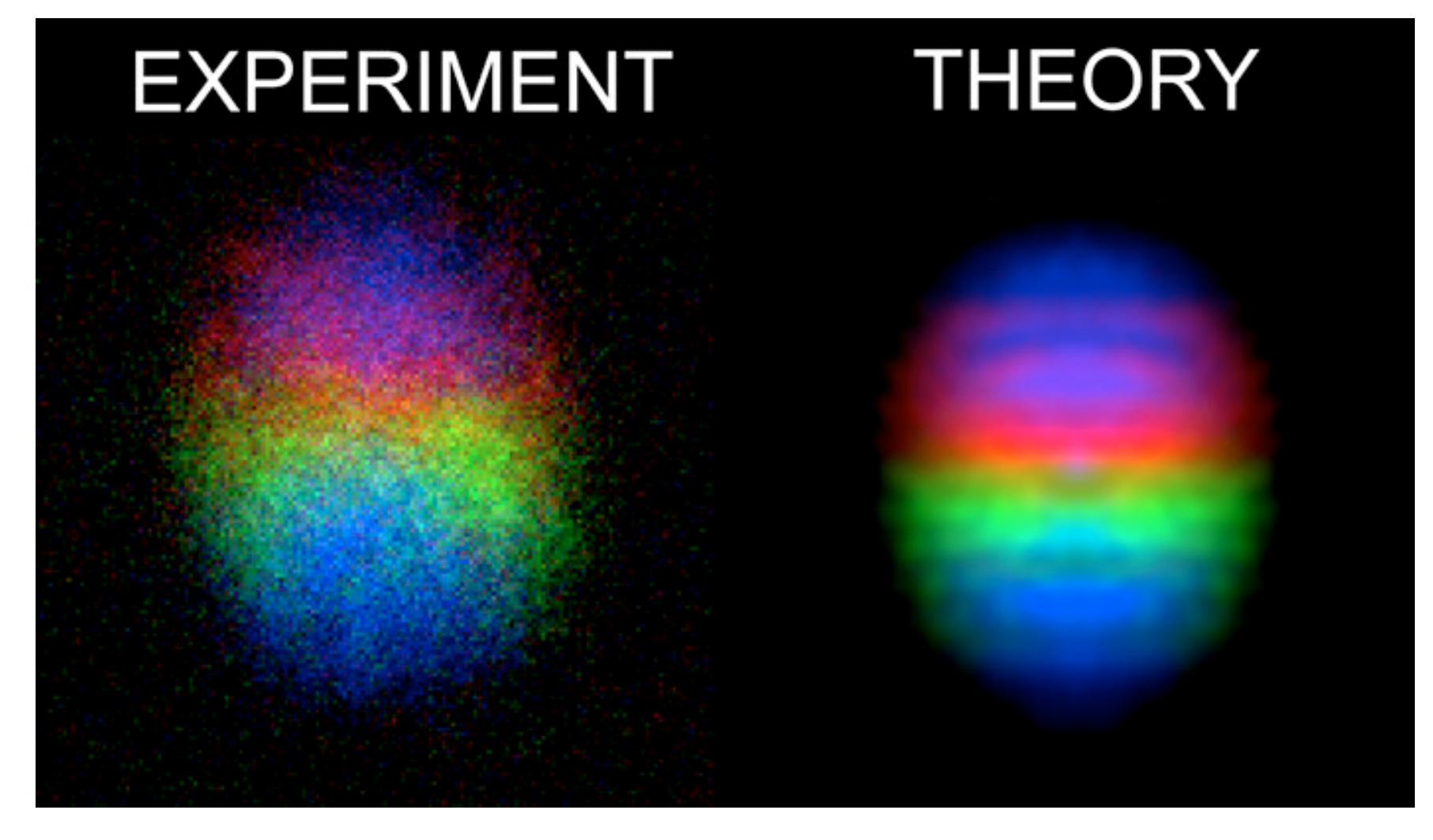}
\caption{Experimental observation and theoretical prediction of a monopole in BEC. Reproduced from \cite{bec_monopole}. Creative Commons Attribution License (CC BY) \protect\url{https://creativecommons.org/licenses/by/4.0/}}
\label{fig11}
\end{figure}

\jcmindex{M}{Monopole}
As such, free magnetic monopoles do not exist in nature but recent advances in condensed matter and atomic physics have demonstrated the existence of effective magnetic
monopoles in artificial spin ice \cite{phatak, nisoli}, chiral magnets \cite{tube} and BECs \cite{monopole_bec, bec_monopole}.  The latter are created in $^{87}$Rb atom condensates
in a synthetic magnetic field. Note that a monopole-antimonopole pair is necessarily connected
by a Dirac string. When two skyrmion tubes touch at a point it creates an effective magnetic monopole because the emergent magnetic field [see Eq. (\ref{eq3b})] at that point is radially
outward \cite{tube}, see Fig. \ref{fig9}.  Similarly, a moving heldgehog at the end of a skyrmion line (in a ferromagnetic nanowire) constitutes an emergent magnetic monopole \cite{monopole},
see Fig. \ref{fig10}.  An experimentally observed and simulated monopole in a Bose-Einstein condensate is depicted in Fig. \ref{fig11}.

\section{Vortices and Vortex Loops}\jcmindex{V}{Vortex} \jcmindext{V}{Vortex}{loop}
In many physical systems such as fluids, superconductors and magnets
(and those modeled by the 3D Heisenberg model) vortices, vortex lines
and vortex loops
(rings) \cite{loops}
are observed.  Creation and dynamics of trefoil-like (and other) knotted vortices have been studied in water using specially shaped hydrofoils \cite{vortexknot}.  Since magnetic,
superconducting and other vortices as well as their dynamics have been
studied extensively, this is a fully developed area of research and
thus we will not dwell on this
general theme, but rather limit ourselves to a number of recent developments.

Vortices are persistent circulating flow patterns that occur in diverse
scientific contexts \cite{Pismen1999}, ranging from
hydrodynamics, superfluids, and nonlinear optics~\cite{Kivshar-LutherDavies,YSKPiO}
to specific instantiations in sunspots~\cite{sunspots}, dust devils~\cite{Lugt1983}, and plant
propulsion~\cite{Whitaker2010}. The study of the associated
2D effective
particle dynamics that results from the logarithmic interaction potential
is a theme of broad interest in physics. Not only it is relevant
for the prototypical fluid/superfluid applications (see
e.g.~the review of Aref et al.~\cite{aref1} and the book of Newton~\cite{newton1}),
but also for a variety of other settings. As such, we mention
electron columns in Malmberg-Penning
traps~\cite{fajans} and magnetized, millimeter sized
disks rotating at a liquid-air interface~\cite{whitesides,whitesides2},
among others.

\jcmindex{B}{Bose-Einstein condensate}
The realm of atomic BECs~\cite{review_dalfovo,becbook1,becbook2} has produced a novel
and pristine setting where numerous features of the exciting
nonlinear dynamics of single- and multi-charge vortices, as well as of
vortex lattices, can be not only theoretically
studied, but also experimentally observed.
Although BEC is known to be a fundamental phenomenon
connected, e.g. to superfluidity and superconductivity
\cite{chap01:becgss},
BECs were only experimentally realized 70 years later:
this major achievement took place in 1995 \cite{chap01:anderson,chap01:davis,chap01:bradley} and has already been recognized
through the 2001 Nobel prize in Physics \cite{chap01:rmpnlA,chap01:rmpnlB}.
The role of vortices and the remarkable manifestation of highly ordered,
triangular vortex lattices were, in turn, cited in the 2003 Nobel
Prize in Physics~\cite{needcite}.
Importantly, vortex dipoles (pairs of oppositely charged vortices)
that will be relevant in what
follows have played a quintessential role in the Kosterlitz-Thouless (KT)
transition~\cite{needcite2} from a gas of dipoles to configurations of unbound
vortices, earning its discoverers the 2016 Nobel Prize in Physics!
This transition has, moreover, found one of its most canonical
realizations in the context of atomic BECs~\cite{dalikruger}.

In addition to being at the epicenter of
some of the most important physical notions of the past few decades,
the coherent structures considered herein have been recognized as
having both practical, as well as more exotic applications.
For instance,
solitary waves have been argued to provide the potential
for 100-fold improved sensitivity for interferometers to
phase shifts~\cite{negretti}, while their lifetime
of a few seconds enables precise force sensing applications~\cite{markk}.
Moreover, vortices present their own potential
for applications. An intriguing example is the so-called
``analogue gravity'', whereby they may play a role similar
to spinning black holes. This allows to observe in terrestrial, experimentally
controllable environments associated phenomena such as
the celebrated Hawking radiation or simpler ones such as super-radiant
amplification of sonic waves scattered from  black holes~\cite{savage}.
It has also been recently argued that vortices of a
rotating BEC can collapse
towards the generation of supermassive black holes~\cite{dasgupta}
and that supersonically expanding BECs can emulate properties of
an expanding universe in the lab~\cite{gretchen}.

The first experimental observation of BEC vortices~\cite{Matthews99}
paved the way for a systematic
investigation of the dynamical properties of such entities.
Stirring the BECs \cite{Madison00,Williams99} above a certain critical angular speed
\cite{Recati01,Sinha01,Madison01} led to the production of few vortices
\cite{Madison01} and
vortex lattices \cite{Raman,jamil}. Other vortex-generation
techniques were also
used in experiments, including the breakup of the BEC superfluidity by
dragging obstacles through the condensate \cite{kett99}, as well as
nonlinear interference between condensate fragments \cite{BPAPRL}.
In addition, apart from
unit-charged vortices,
higher-charged vortex structures were produced \cite{S2Ket1,S2Ket2} and their dynamical
(in)stability was examined.

\begin{figure}[tb]
\vskip-0.3cm
\psfig{file=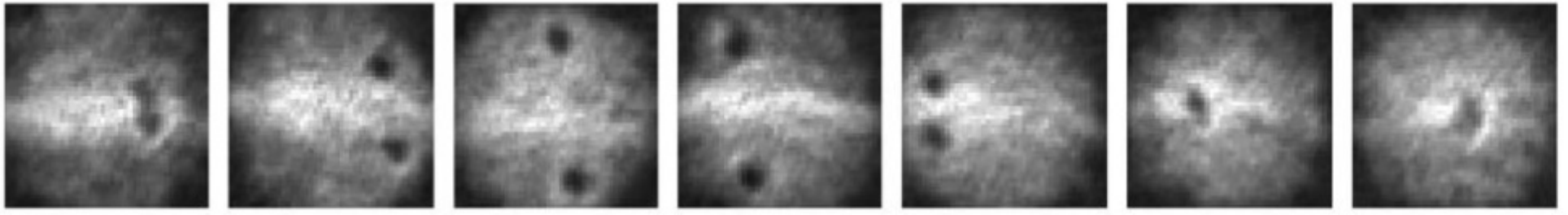,height=1.400cm,angle=0,silent=}
\hskip-0.1cm
\psfig{file=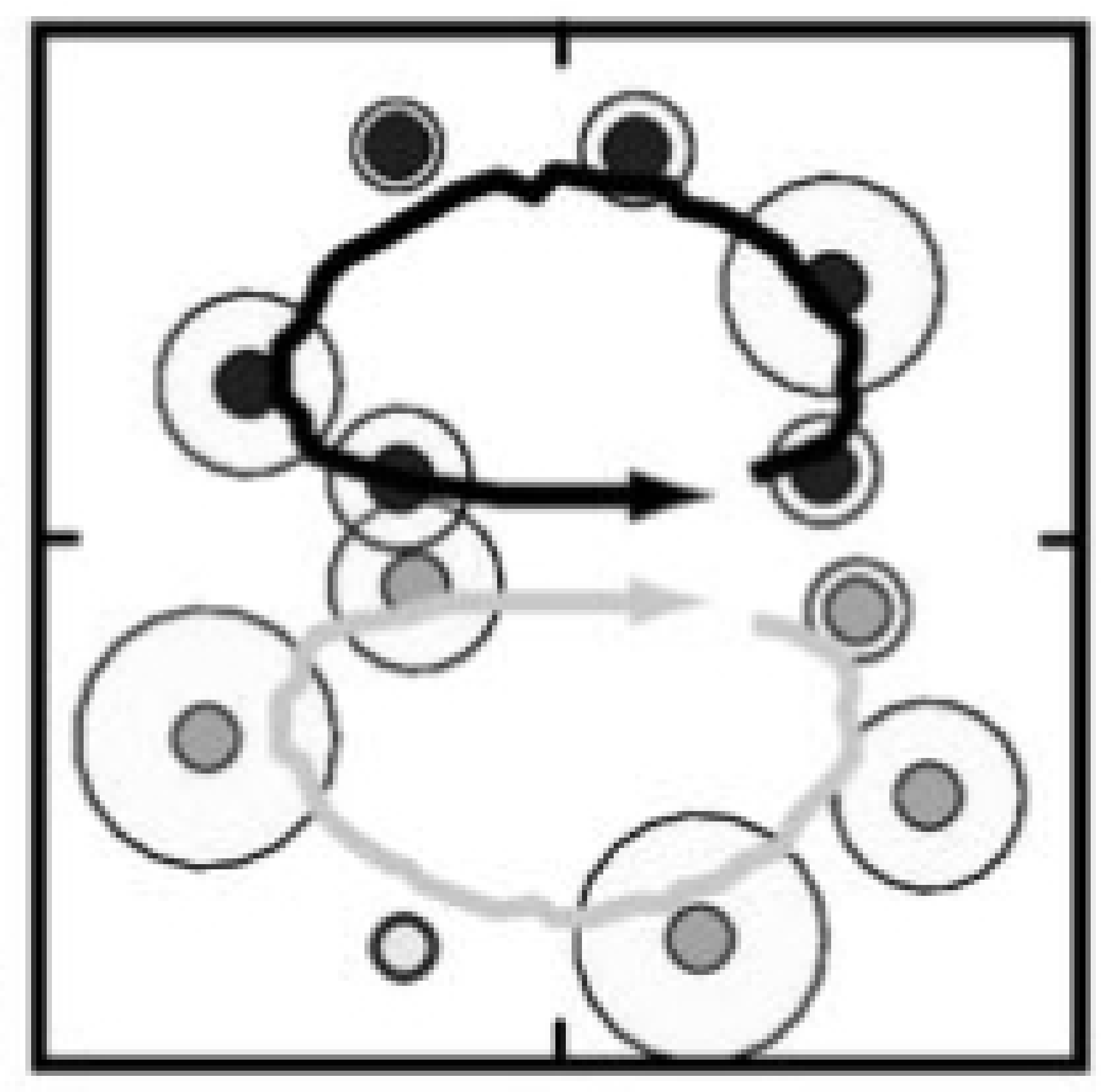,height=1.400cm,angle=0,silent=}\\
\psfig{file=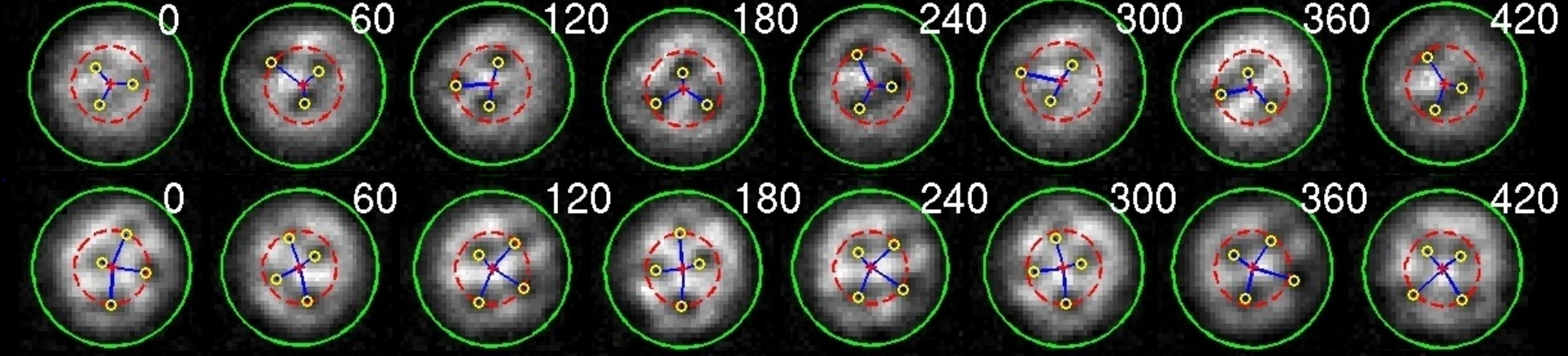,height=1.996cm,angle=0,silent=}
\caption[]{
Sampler of small vortex clusters in recent BEC experiments. Top row: vortex dipole dynamics; reproduced from \cite{BPA3} (\copyright 2010 by the Americal Physical Society). Bottom row: dynamics for
three (top row) and four (bottom row) same-charge vortices reported in~\cite{dshall3}}
\label{fign1}
\end{figure}

\jcmindex{K}{Kibble-Zurek mechanism}
The majority of these early experiments focused on creating individual vortices and
large vortex arrays. However, in 2008, the work of~\cite{BPA} enabled the use of the so-called Kibble-Zurek (KZ) mechanism
to quench a gas of atoms rapidly across the BEC transition. The result of this is
that phase gradients do not have sufficient time to ``heal'' but rather freeze,
resulting in the formation of vortices.
Then, in 2010 another technique was devised
that enabled for the first time the
dynamical visualization of
vortices~\cite{dshall} during an experiment.
This, in turn, spearheaded the work of~\cite{dshall1,dshall2}
where particle models were developed that predicted the dipole dynamics
(equilibria, near-equilibrium epicyclic precessions and far from
equilibrium quasi-periodic motions) observed in these experiments.
A nearly concurrent development concerned the production in the lab
of such vortex dipoles (one or multiple such),
by the superfluid analogue of dragging a cylinder through a fluid~\cite{BPA3}.
More recently,
the KZ mechanism together with
rotation (i.e. injection of angular momentum) have been used
to ``dial in'' and observe the dynamics of
vortex clusters of, controllably, any number of vortices between 1 and 11.
This is because rotation favors the formation of vortices
of the same charge and in this way, depending on the angular momentum
provided, different charge configurations (vortex clusters) arise.
The resulting configurations may suffer symmetry breaking
events~\cite{dshall3}.
As a result,
instead of the commonly expected anti-diametric pair,
equilateral triangle, or square
configurations that one may expect, it is possible to observe
symmetry broken configurations featuring asymmetric pairs,
isosceles triangles, and rhombi or general/asymmetric
quadrilaterals~\cite{zampetaki}. To further add to these developments, 3-vortex configurations of (same but also
of) alternating charge in the form of a tripole (i.e. a positive-negative-positive
or its opposite) have been experimentally produced~\cite{tripole,vd_v14}.
This turns out to be one of the simplest setups
where chaotic dynamics can ensue~\cite{vkouk1,vkouk2}.
A sampler of experimental images from these different experimental efforts is
depicted in Fig.~\ref{fign1}.

It is evident from the above recent developments that there is
a tremendous momentum toward the study
of vortex dynamics in
atomic physics. Moreover, this
theme presents nontrivial twists in comparison to the classical
fluid or superfluid playground~\cite{aref1,newton1}. To mention a canonical
difference between the two, atomic BECs are
typically confined by parabolic
traps~\cite{review_dalfovo,becbook1,becbook2,emergent}
constraining the density
and hence the region within which the vortices evolve.
This trapping induces vortices to rotate around the center of
the trap~\cite{emergent,fetter1,fetter2}. The frequency of this
precessional motion
can be well approximated by a constant close to the center
of the trap, yet as the edges of the BEC are approached and the density
decreases, the relevant frequency increases drastically~\cite{dshall3,zampetaki,theo14}.
These modifications of the ``standard'' picture of the
vortices interacting through a logarithmic potential are critically
responsible for some of the phenomena observed recently. For instance,
the competition between the rotation and the interaction in the case
of the vortex dipole~\cite{dshall1,dshall2} is
responsible for the existence of stationary states
or epicyclic/quasi-periodic trajectories; the deviation
from a constant precession frequency is, in turn,
responsible~\cite{zampetaki,theo14} for the symmetry-breaking
bifurcations enabling asymmetric vortex configurations~\cite{dshall3}.

\section{Skyrmions in Liquid Crystals} \jcmindex{L}{Liquid crystals}
\label{sec:liq}
One can describe a liquid crystal by a tensor order parameter
$\mathbf{Q}(\mathbf{r})$, which is related to  the director field
$\mathbf{n}(\mathbf{r})$ and the scalar order
parameter $S(\mathbf{r})$  by
$Q_{\alpha\beta}=S(\frac{3}{2}n_\alpha n_\beta
-\frac{1}{2}\delta_{\alpha\beta})$.  According to the Landau-de Gennes theory, one can express the free energy density in terms of $\mathbf{Q}$ as \cite{liquidcrystal}
\begin{eqnarray}
\label{freeenergy}
F&=&\frac{1}{2}a Tr\mathbf{Q}^2+\frac{1}{3}b Tr\mathbf{Q}^3+\frac{1}{4}c\left(Tr\mathbf{Q}^2\right)^2\\
&&+\frac{1}{2}L(\partial_{\gamma}Q_{\alpha\beta})(\partial_{\gamma}Q_{\alpha\beta})- 2Lq_0 \epsilon_{\alpha\beta\gamma}Q_{\alpha\delta} \partial_{\gamma}Q_{\beta\delta}.\nonumber
\end{eqnarray}
The first line represents the free energy of a uniform system, when we expand it in powers of the tensor order parameter.  This part favors those eigenvalues of $\mathbf{Q}$, which are associated with a specific magnitude of uniaxial nematic order.  One assumes the coefficient $a$ varies linearly with temperature, whereas $b$ and $c$ are constant with regard to temperature.  The last two terms are the elastic free energy corresponding to variations in $\mathbf{Q}$ as a function of position. The first of these terms is the (equal) energy cost of splay, twist, and bend deformations, where $L$ is an elastic coefficient.  The last term allows a chiral twist of the nematic order, where $q_0$ is a characteristic inverse length that arises from the molecular chirality.  Other possible elastic terms giving different energy costs for splay, twist, and bend, e.g. $\frac{1}{2}L_2  (\partial_{\alpha}Q_{\alpha\gamma})(\partial_{\beta}Q_{\beta\gamma})$, are neglected here for simplicity.

The chiral twist term $q_0$ is analogous to the Dzyaloshinskii-Moriya interaction in the magnetic case [see Eq. (\ref{eq2})]. Comparative analysis and simulations based on (\ref{eq2}) for the magnetic case and (\ref{freeenergy}) for the liquid crystals \cite{liquidcrystal} provide the results depicted in Fig. \ref{fig3}.  For chiral magnets the triangular (or hexagonal) skyrmion lattice is stable
and it arises from the spiral (or helical) phase.  However, a
triangular meron lattice is not stable but a square meron lattice is
allowed.
In contrast, based on energetic grounds, in liquid crystals the skyrmion lattice is disfavored but a triangular lattice of merons is allowed. This
difference arises from the nature of the order parameter: vector for chiral magnets versus tensor for nematic liquid crystals.

In addition to skyrmions and merons, stable skyrmion bags have been observed in liquid crystals \cite{bags}.  An observed and simulated example is illustrated in Fig. \ref{fig5}.  Corresponding skyrmions bags in chiral magnets are also possible as shown in Fig. \ref{fig5} as a result of micromagnetic simulations.  Interestingly, these bags are similar to the models of atomic nuclei containing different number of baryons, as originally
surmised by T.H.R. Skyrme.

\section{Bose-Einstein condensates: From Vortex Lines to Rings, From Hopfions to Skyrmions and Knots}
\label{sec:bec}

We already discussed in Sec. 4 the relevance
of Bose-Einstein condensates as a prototypical playground where
two-dimensional topological excitations in the form of vortices naturally arise.
We now turn to a 3D extension of such structures, starting with the
natural
generalization of the vortex, namely the vortex line (VL). \jcmindext{V}{Vortex}{line}
A VL, also referred to as a solitonic vortex,
is the 3D extension of a 2D vortex by (infinitely and homogeneously)
extending the solution into the axis perpendicular to the vortex plane.
VLs might be rendered finite in length if their background
is made bounded by an external
potential.
In that case, VLs are called vorticity ``tubes'' that are straight
or bent in U and S shapes depending on the aspect ratio of
the background \cite{VR:USshapedVLs1,VR:USshapedVLs2}.
If a VL is bent enough to close on to itself or if two
VLs are close enough to each other then they can produce
a vortex ring \cite{VR:CrowInstab}.
Vortex rings (VRs) \jcmindext{V}{Vortex}{ring} are 3D structures whose core is a closed loop with vorticity
around it \cite{donnelly} (i.e. a vortex that is looped back into itself).
VRs can also be produced by an impurity traveling faster
than the speed of sound of the background \cite{rcg:62N},
by nonlinear interference between colliding blobs
of atomic matter~\cite{rcg:55,ourbrian1},
by phase and density engineering techniques
\cite{VR:Dutton1,chap01:dark,SteinhauerNatPhys},
or even by introducing ``bubbles'' of one component in the other
component in two-component nonlinear Schr\"odinger (NLS) systems \cite{VR:2C}.

\begin{figure}[t]
\sidecaption[t]
\includegraphics[width=7cm]{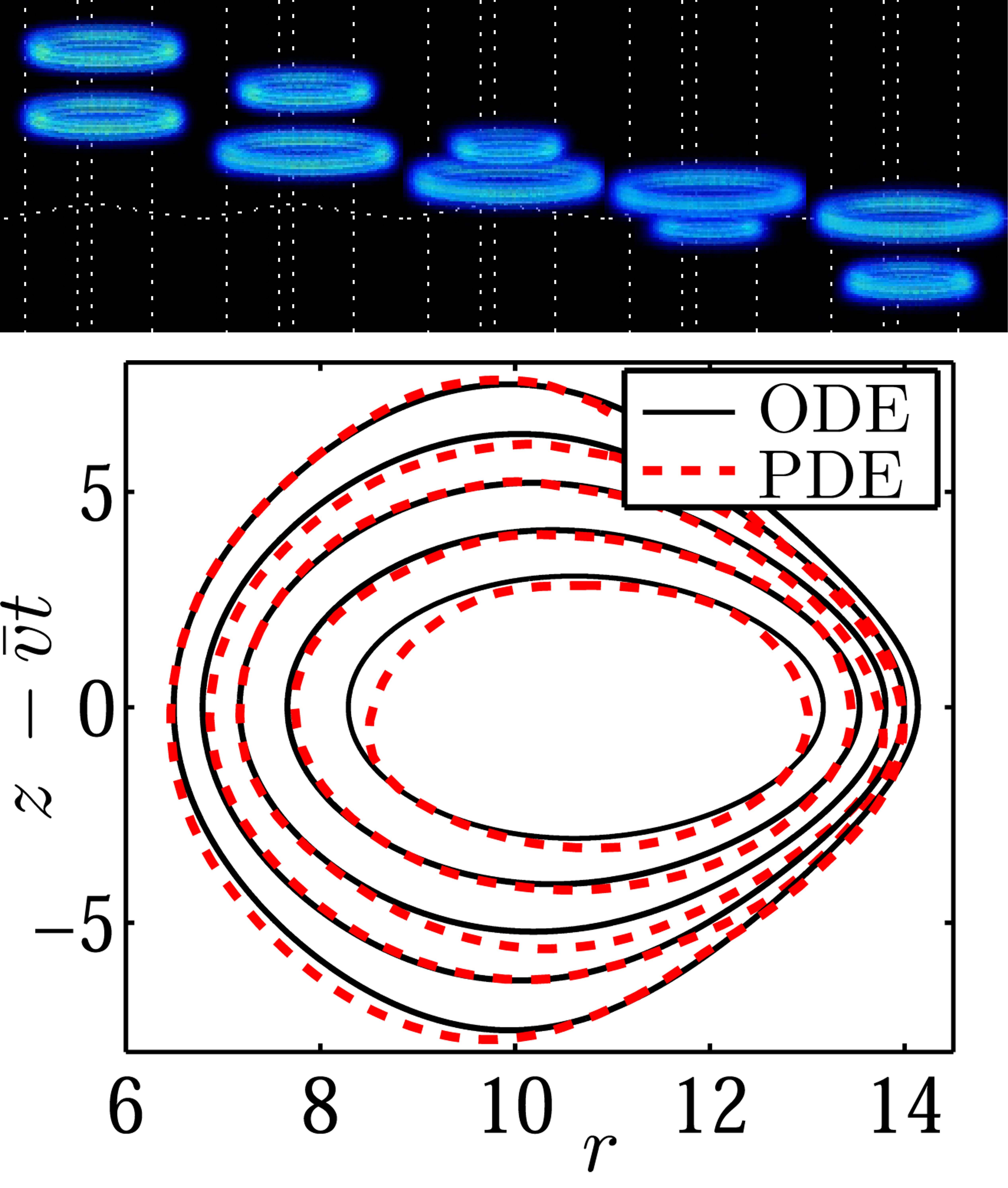}
\caption{The top panel shows the snapshots of the evolution of
  two vortex rings leapfrogging through each other. The bottom panel
  illustrates this type of motion in the radial-polar $(r,z)$ plane on a co-moving
reference frame. The solid lines correspond to the
orbits predicted by ordinary differential equations for the radius and vertical
position of the rings as a function of time.
The dashed lines are the corresponding numerics of the full
3-dimensional NLS equation. Bottom panel is adapted from
\cite{Caplan}. \copyright 2016 by the American Institute of Physics}
\label{f:plotRZ}
\end{figure}

It should be noted that VRs inherently
possess a velocity perpendicular to the ring plane due to Helmholtz's
law \cite{vrvel71} (unless they are stopped by the
presence of an external trap \cite{VR:Guilleumas1,VR:Guilleumas2}).
Also, another special feature of VLs and VRs is that they support
intrinsic dynamics along the vortex line/ring. For example,
it is possible to transversally excite the vorticity line to
produce oscillations called Kelvin modes (or Kelvons)
\cite{VR:KelvinModesFetter,VR:KelvinVaricose,VR:Taiwanese,kelvin:stringari}.
Kelvin modes not only have their own dynamics and interactions
across vortex lines \cite{VR:VictorPG1}, but they can also
self interact within a single VR and slowdown or even reverse
the velocity of the VR \cite{VR:SlowingDown1,VR:SlowingDown2}.
We note here that Kelvin modes have also been studied in the context of
skyrmion tubes \cite{skykelvin}.
Another possibility for exciting the vorticity line of the VR is by
creating varicose or capillary waves (periodic compressions of the vortex
tube along its length) \cite{VR:KelvinVaricose,VR:Capillary}.
Lastly, VRs interact in intriguing ways involving, e.g.
leapfrogging motions when they are co-axial
(see Fig.~\ref{f:plotRZ}), but also
more complex interactions when they are not~\cite{saffman}.
Recently, an effective particle description has been utilized
not only in order to understand the stability and dynamics
of a single VR~\cite{usVR1,usVR2}, but also that of multiple
or interacting VRs~\cite{usVR3}.

It is important to highlight here that the relevance of VLs and VRs
goes beyond atomic BECs. They emerge ubiquitously
in fluid mechanics~\cite{saffman} and in Helium and other related
superfluid systems~\cite{donnelly}.
Rather, what is the case here
is that atomic condensates present a pristine, well-controlled
setting for the creation and exploration of these structures.

In addition to VRs and VLs, it
was also realized that BECs offer also the potential for
the formation of more
complex topological structures.
This is to a considerable extent due to the potential of creating atomic
condensates either of different species (e.g. $^{87}$Rb and
$^{23}$Na, i.e. hetero-nuclear mixtures) or of the same
species (e.g. confining and condensing two different
hyperfine states of the same gas, such as spin-1 and
spin-2 states of $^{87}$Rb)~\cite{emergent}.
Among the early suggestions along
this vein, is the multi-component  Skyrmion state creation in BECs.
The
topological properties of such a state enable its structural
realization in a multi-component BEC.
Here, as originally proposed
in \cite{Ruostekoski01,ruost}, the Skyrmion consists of a VR in one
of the components,
``trapping'' a VL in a second component. Interestingly, more
complex Skyrmion states involving three-component (so-called spinor)
BECs have been recently realized experimentally in both 2D~\cite{shin} and
3D~\cite{bigelow},
involving, respectively, coupled states of topological charge
$S=-1, 0, 1$ and $S=0, 1, 2$; see also the recent
theoretical work of \cite{ruost2}.

Of increased interest recently
has been not only this structure, but also its one-component
counterpart, which in the relevant recent BEC literature
is referred to as a hopfion state~\cite{boris,yakimenko,ushop}.
The latter consists of a VR and a VL in the same component with the
axis of the VR constituting the line of vorticity of the VL.
A stable hopfion state was found to exist both in the setup of \cite{boris}
which, however, involved the rather elaborate realization of radially
increasing nonlinear interactions and the purely dynamical exploration
of \cite{yakimenko} for condensates that are being rotated (and was
suggested to be stable only for some
intermediate rotation rate).

\jcmindex{K}{Knot}
Lastly, the study of quantum (vortex) knots is one that has only a relatively
short history in the context of atomic BECs. To the best of our knowledge,
the possibility of such complex topological structures was
introduced in the work of~\cite{barenghi12} (see also~\cite{bar14}),
illustrating how different torus knots ${\cal T}_{K,q}$,
with co-prime $K$ and $q$,
can be generated in the wavefunction of an atomic species.
Subsequently the work of~\cite{irvine} seemed to put a full stop
on the subject through the extensive simulation of 1458 vortex
knots from the so-called ``knot atlas''~\cite{knotatlas}, and finding that the
{\em trapless} Gross-Pitaevskii equation (GPE) could not support stable knots:
all of the simulated knots would eventually untie into simpler patterns.
Nevertheless, the recent work
of~\cite{ruban3,ruban4}, both at the level of the Biot-Savart dynamical
law (for the vortex knot motion), as well as at that of the full
3D GPE has, perhaps counter-intuitively, indicated that the {\em trefoil
knot can be a very long lived} structure in the context of a trapped atomic
BEC~\cite{usknot}.

\section{Topology and curved manifolds: Bogomolnyi decomposition}
\label{sec:curved}

We now provide an example of how the types of configurations
considered
herein may be used to minimize the energy of a system of classical spins.
The continuum limit of (classical) Heisenberg spins on a two-dimensional (planar or) curved manifold corresponds to the nonlinear $\sigma$ model. \jcmindex{N}{Nonlinear $\sigma$ model} {That is, the corresponding Hamiltonian $H$ is given by (\ref{eq:sigma}).  If we impose
homogeneous boundary conditions on the vector field ${\mathbf n}$ in the plane $R^2$, i.e. $\lim_{r\rightarrow\infty} {\mathbf n} \rightarrow {\mathbf n}_0$, then
we can compactify the plane into the surface of a sphere $S^2$.  This allows us to classify different configurations according to the homotopy class
$\pi_2(S^2) = {\mathbf Z}$, where ${\mathbf Z}$ is the group of relative integers \cite{cylinder, BP}.

Topology does not directly help us to state anything about the energy
of the field configuration but indirectly, by invoking the so-called
Bogomolnyi inequalities \cite{bogom},
it enables us to establish energy bounds for configurations belonging to equivalent homotopy classes labeled by $n \in {\mathbf Z}$. The inequality in the
present case can be expressed as
\begin{equation}
( \partial_i{\mathbf n} - \epsilon_{ij}\partial_j{\mathbf n})^2 \ge 0 \,,
\end{equation}
whereby it follows that
\begin{equation}
H \ge \int {\mathbf n}\cdot (\partial_x{\mathbf n}\times \partial_y{\mathbf n}) dxdy \,.
\end{equation}
Thus, the minimum energy in each homotopy class is attained when
\begin{equation}
\partial_i{\mathbf n} = \pm \epsilon_{ij} \partial_j{\mathbf n} \,,
\end{equation}
i.e. when these self-dual equations are satisfied by the field configurations.

If we consider this model on a plane ($R^2$), there is no characteristic length.  As a result the nonlinear $\sigma$ model Hamiltonian can be scaled and
thus all the nontrivial field configurations (satisfying homogeneous boundary conditions)  can be scaled as well.  This situation is drastically changed if
there is a characteristic length scale, e.g. if the underlying manifold is curved.   We will introduce a length scale in two ways: (i) first we will consider the
nonlinear $\sigma$ model on a rigid cylinder and then (ii) we will also apply an axial magnetic field through the cylinder.  In the first case the radius $\rho_0$
of the cylinder is the characteristic length whereas in the second case there is an {\it additional} length scale introduced by the magnetic field ${\bf B}$.
There are other ways of introducing a length scale, e.g. through magnetic anisotropy, ellipticity of the cylinder cross section, etc. but we will not consider these
different cases here.

We set the unit vector ${\bf n} = (\sin\theta \cos\Phi, \sin\theta \sin\Phi, \cos\theta)$ in terms of the co-latitude $\theta$ and the azimuthal angle $\Phi$.
Next, we write the Hamiltonian \cite{cylinder} in terms of cylindrical
coordinates $(\rho, z, \phi)$
\begin{equation}
H = J \int\!\!\!\int_{\mathrm{cyl}} \left[(\partial_z\theta)^2 + \sin^2\theta (\partial_z\Phi)^2 +(\partial_\phi\theta)^2/\rho_0^2 + \sin^2\theta (\partial_\phi\Phi)^2/\rho_0^2 \right]
\rho_0 dz d\phi \,,
\end{equation}
where $J$ denotes the spin exchange interaction energy. In order to invoke topological considerations let us impose homogeneous boundary conditions,
i.e. $\lim_{z\rightarrow\mp\infty} \equiv 0[\pi]$ and $\lim_{z\rightarrow\mp\infty} d\theta/dz=0$.  If we seek cylindrically symmetric solutions then $\Phi=\phi$
and $\partial\theta/\partial\phi=0$.  Thus the Hamiltonian simplfies to
\begin{equation}
H = 2\pi \rho_0 J \int_{-\infty}^\infty \left[(\partial_z\theta)^2 +\sin^2\theta /\rho_0^2 \right] dz \,.
\end{equation}
The variation of this Hamiltonian ($\delta H=0$), i.e. the Euler-Lagrange equation turns out to be the celebrated sine-Gordon equation
\begin{equation}
d^2\theta(z)/dz^2 = (1/2\rho_0^2) \sin2\theta \,,
\end{equation}
with the well known kink solution $\theta(z) = \arctan[\exp(z/\rho_0)]$.  It is depicted in Fig. \ref{fig13}(a). The energy for this configuration is $H=8\pi J$,
which is the minimum energy belonging to the first homotopy class.

\jcmindex{B}{Bogomolnyi decomposition}
By invoking the technique used by Belavin and Polyakov \cite{BP}, or equivalently the Bogomolnyi decomposition \cite{bogom}, we note that the solutions that
correspond to the minimum energy in each homotopy class satisfy the first order self-dual equations
\begin{equation}
\rho_0 \partial_x\theta = \pm \sin\theta \partial_\phi\Phi \,, ~~~  \partial_\phi\theta = \mp \rho_0 \sin\theta \partial_z\Phi \,.
\end{equation}

\begin{figure}[t]
\centering
\includegraphics[width=.9\textwidth]{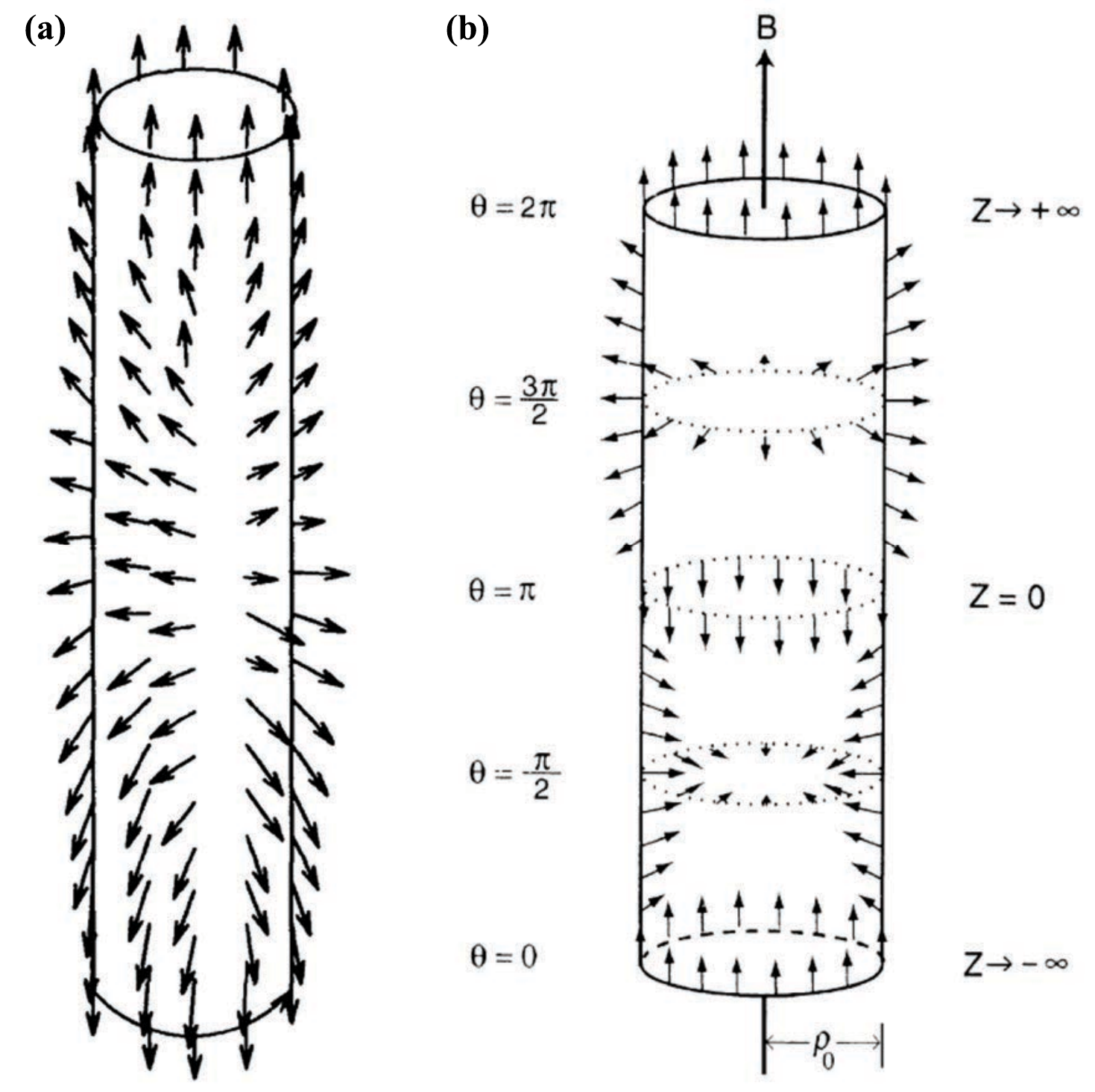}
\caption{(a) Heisenberg spins on a cylinder as a sine-Gordon $\pi$-soliton (reproduced from \cite{bogom2}, \copyright 1994 by Elsevier Science B.V.) and (b) in the presence of an axial magnetic field as a double sine-Gordon $2\pi$ soliton (reproduced from \cite{DSG})}
\label{fig13}
\end{figure}

If we apply an external magnetic field (${\bf B}$) along the axis of the cylinder then the Hamiltonian is modified as \cite{DSG}
\begin{equation}
H_\mathrm{mag} = J \int\!\!\!\int_{\mathrm{cyl}} (\nabla n)^2 dS -g\mu \int\!\!\!\int_{\mathrm{cyl}} {\mathbf n}\cdot{\mathbf B} ~dS \,,
\end{equation}
where $g$ is the gyromagnetic ratio and $\mu$ denotes the magnetic moment.  With homogeneous boundary conditions ($\theta=0$ as $z\rightarrow \pm\infty$)
the Hamiltonian simplifies as
\begin{equation}
H_\mathrm{mag} = 2J(2\pi\rho_0)\int_{-\infty}^\infty [\theta_z^2 + \{\sin^2\theta/2\rho_0^2 + (1/\rho_B^2)(1-\cos\theta)  \} ] dz\,,
\end{equation}
where $\rho_B^2= 2J/g\mu B$ is the magnetic length scale.  Variation of this Hamiltonian leads to
\begin{equation}
\theta_{zz} = (1/2\rho_0^2) \sin2\theta + (1/\rho_B^2)\sin\theta \,,
\end{equation}
which is the double sine-Gordon equation.  The corresponding $2\pi$-kink solution that is consistent with the boundary conditions is given by
\begin{equation}
\theta(z) = 2\arcsin \frac{1}{\sqrt{\cosh^2 (z/\xi)  - (\xi^2/\rho_0^2) \sinh^2(z/\xi)}}   \,,
\end{equation}
where the kink width $\xi = \rho_0\rho_B/(\rho^2+\rho_B^2)^{1/2}$ is another characteristic length in the problem.  This solution is depicted in Fig. \ref{fig13}(b).

Because of the homogeneous boundary conditions at the cylinder boundaries all the spins point in the same direction, see Fig. 13.  Thus they can be compactified
to a single spin and the spin configuration in Fig. 13(a) covers the unit sphere once, i.e. it is a skyrmion of topological charge 1.  Similarly, the spin configuration in Fig. 13(b)
covers the unit sphere twice, thus it is a skyrmion of topological charge 2.  If the cylinder were semi-infinite, it will be topologically equivalent to a plane with a hole of radius
$\rho_0$.  The spin configuration in this case will be a half-skyrmion (or a meron) \cite{halfsky}.

If the cylinder is elastic (i.e. deformable) then the geometric frustration caused by the mismatch of the cylinder radius and kink width can be relieved by a pulse-like
deformation in the region of the magnetic kink \cite{cylinder, DSG}.  The Bogomolnyi technique is quite general and can be used in a broader context. Another application
of the Bogomolnyi decomposition is in the calculation of elastic deformation energy of vesicles as a function of genus \cite{vesicles}. The latter has significance in the
context of the Willmore conjecture \cite{willmore}.


\section{Topological Materials}

In conventional materials such as metals, insulators and semiconductors the non-relativistic Schr\"odinger equation describes the energy dispersion of low-lying electronic excitations,
$E_S=p^2/2m^*$, which is quadratic in the electron momentum $p$ with effective mass $m^*$. However, over the past decade there is a growing class of materials, which are known as
Dirac materials \cite{wehling} or more generally topological materials, exhibiting linear electronic dispersion in their band structure.  Examples include topological insulators \cite{TI, TI2}, topological superconductors \cite{topsup}, topological crystalline insulators \cite{TCI} as well as Dirac semimetals and Weyl semimetals \cite{semimetals,vafek}.
\jcmindex{D}{Dirac materials}

One of the distinguishing features of these materials are Dirac
points, where the (conduction and valence) bands touch each other at
an isolated set of points.  The corresponding band features are called
Dirac cones.  These points are topologically protected due to specific
(time reversal, spatial inversion or crystalline) symmetries in that
they are robust under perturbations. An example is graphene where the
protection comes from the sublattice symmetry (of the underlying
honeycomb lattice) and the energy dispersion of its electrons is
linear in the momentum.  Specifically, it is given by the relativistic Dirac equation: $E_D = c \sigma\cdot p + mc^2 \sigma$.  Here $\sigma=(\sigma_x, \sigma_y)$ denotes Pauli matrices and the speed of light $c$ is replaced by the Fermi velocity $v_F$.  In $d$ spatial dimensions (with $c=1$) the Dirac equation is written as $(i\gamma^\mu\partial_\mu-m)\psi=0$
\cite{semimetals}, where $\mu=0,1,..., d$ with $\mu=0$ denoting time and the Dirac gamma matrices $\gamma^\mu$ anticommute.  In odd dimensions ($d=1,3, ...$) it can be simplified.  In particular, for $d=1$ one gets  $i\partial_t\psi=(\gamma^0\gamma^1 p + m\gamma^0)\psi$ with momentum $p=-i\partial_x$.  If we further consider massless ($m=0$) particles, we get the one-dimensional Weyl equation $i\partial_t\psi_{\pm} = \pm p\psi_{\pm}$.  Thus we get simple linear dispersion $E_\pm = \pm p$ representing the right and left moving chiral particles or Weyl fermions.

Topology seems to enhance the nonlinear response of topological materials.  A recent important experimental technique is TFISH (Terahertz Field-Induced Second Harmonic Generation) which allows to understand the nonlinear response of topological Dirac and Weyl semimetals, e.g. TaAs \cite{sirica}. In particular, second harmonic generation
(SHG) is found to be enhanced in Weyl semimetals.  The latter contain Weyl points (or nodes) in their electronic structure at which linearly dispersing, nondegenerate bands cross.
They also exhibit Fermi arc surface states that are attached to the
Weyl nodes in the bulk material (Fig. \ref{fig14}).
In these materials there is a transition between the topological and non-topological (or trivial) phases which proceeds via a gapless state.

\jcmindex{N}{Nonlinear Dirac equation} \jcmindex{N}{Nonlinear Weyl equation}
When an appropriate nonlinearity is added to the usual (linear) Dirac equation and Weyl equation, a nonlinear Dirac (NLD) equation \cite{NLD} and a nonlinear Weyl (NLW) equation
\cite{NLW} results, respectively.  Their properties and attendant nonlinear (topological) excitations such as solitons and vortices \cite{nldvortex} are quite different as compared to
the nonlinear Schr\"odinger (NLS) equation.
Currently, a significant amount of analytical and numerical effort is being devoted to understanding the stability and collision dynamics of solitons in NLD and NLW equations \cite{NLDreview}.

In closing this section we note that in addition to chiral magnets, liquid crystals and BECs, there are many other materials including topological materials in which a variety of topological defects can form under right conditions.  In particular,
ferroelectrics \cite{das}, multiferroic materials, e.g. Cu$_2$OSeO$_3$ \cite{multif} and magnetic shape memory alloys such as Ni$_2$MnGa \cite{ msm} can also support skyrmion-like topological excitations.  There has been a recent observation of skyrmions in the heterostructures of a ferromagnet (Cr$_2$Te$_3$) and a topological insulator (Bi$_2$Te$_3$) \cite{skyTI}.  It would be highly
desirable to observe hopfions in such materials as well.

Although we have not discussed dislocations separately here, they play an important role in determining the properties of topological materials \cite{slager1}.

\begin{figure}[t]
\sidecaption[t]
\includegraphics[width=6cm]{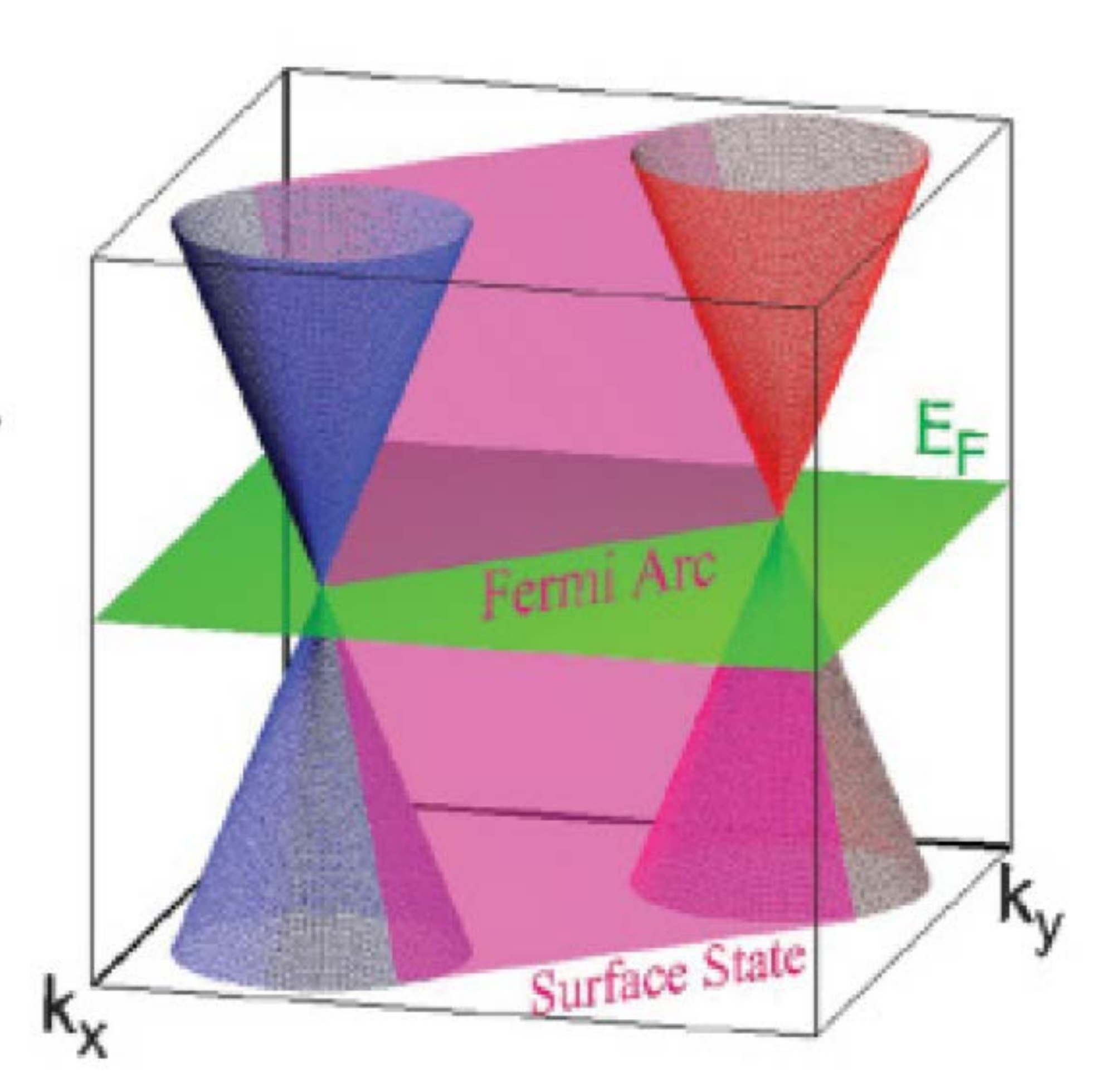}
\caption{The double Dirac cone structure of a Weyl semimetal with two Weyl nodes connected by a Fermi arc.  Here $E_F$ denotes the Fermi energy.  These materials exhibit
characteristic surface states which can be observed experimentally.
Reproduced from \cite{vafek2}. \copyright 2011 by the American Physical Society.}
\label{fig14}
\end{figure}

\section{Nonreciprocal Topological Photonics}

Many of the modern photonic devices such as optical isolators and
optical circulators are based on the principle of Lorentz reciprocity.
It entails that in a (i) linear, (ii) time-independent material or
medium with (iii) symmetric property (or constitutive optical) tensors,
the received and transmitted fields are identical for both forward and
time-reversed propagation directions \cite{photonics}.  However, in
some cases reciprocity is deleterious, as, e.g. in self-echo in antennas.  In addition, for many desired and emerging optical functionalities
(e.g. optical circulators)  it is important to break Lorentz
reciprocity by relaxing any of the  three conditions.  Clearly, one
way to obtain nonreciprocity is by way of introducing optical
nonlinearity.  The latter in conjunction with non-Hermitian photonics
and {\it topological photonics} can significantly enhance
nonreciprocity \cite{nonreciprocity}.  Examples of optical
nonlinearities include the Kerr effect, two-photon absorption and the thermo-optic effect.  An example of nonreciprocal topological photonic setup employing a nonlinear coupled resonator array
is depicted in Fig. \ref{fig15}.

One can create topologically nontrivial photonic band structures in
analogy with the electronic band structure of topological materials
discussed above.  In particular, one can create topological edge
states that are robust against perturbations or defects. One way to
create such states is by forming an interface between a topologically
nontrivial and a trivial optical material. Just like in electronic
topological materials, photonic nontrivial topological bands cannot be
deformed to trivial bands in an adiabatic way.  A photonic realization
of a two-dimensional electronic Chern insulator uses a lattice of magnetized ferrite rods at microwave scale \cite{soljacic} in which the lattice edge state acts as an isolating waveguide.  This structure leads to almost perfect forward transmission and exponentially suppressed backward transmission for frequencies in   the photonic bandgap.

There are two ways of realizing nonlinear topological photonic structures.  One can consider nonlinear propagation dynamics in an otherwise linear topological photonic system.
Here the nonlinearity locally alters the system properties.  An example is that of waveguide arrays with evanescent coupling between neighboring waveguides.   Another way is to
use a probe beam to induce a phase transition in the dynamics of a linearized probe beam.  Coupled optical resonator lattices provide a realization of this type where one gets quite
strong nonlinear effects due to the resonant light confinement as compared to waveguide lattices.  Finally, we note that a one dimensional nontrivial topological lattice can be modeled by the photonic
Su-Schrieffer-Heeger (SSH) model \cite{ssh, alu}.

\begin{figure}[t]
\sidecaption[t]
\includegraphics[width=7.5cm]{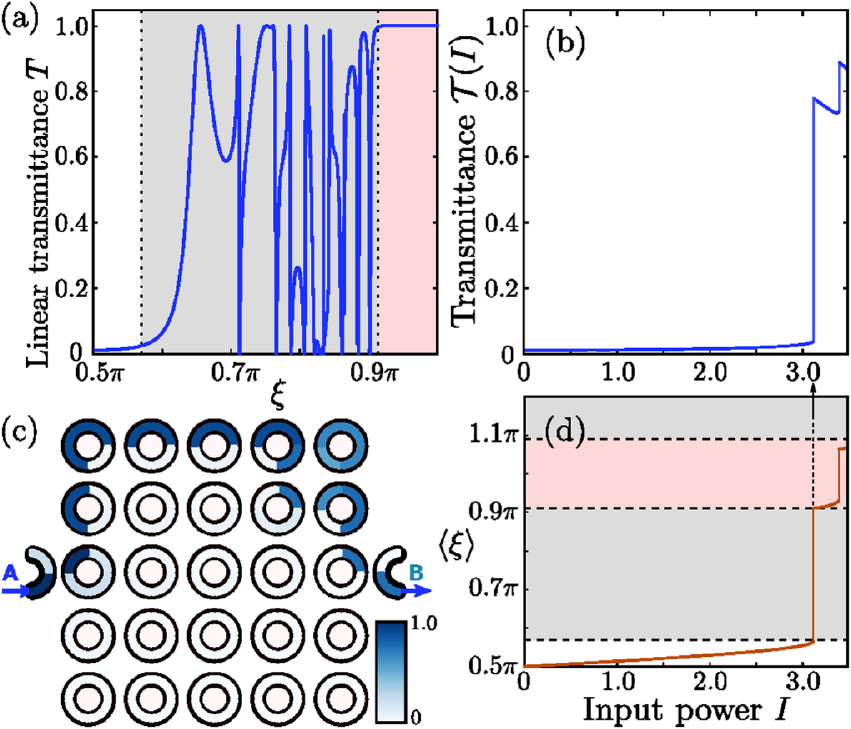}
\caption{Nonlinear coupled ring lattice or resonator array for transmittance studies related to nonreciprocal topological photonics.  (a) Transmittance $T$ through a linear lattice as a function of phase shift $\xi$.  (b) Transmittance as a function of input power $I$ through the nonlinear lattice.  (c) Field intensity distribution (normalized) above the discontinuity ($I>3$). (d) Average phase shift as a function of $I$. Reproduced from \cite{topophotonics}.  Creative Commons Attribution License (CC BY) \protect\url{https://creativecommons.org/licenses/by/3.0/}}
\label{fig15}
\end{figure}


\section{Topological modes in Acoustics and Beyond}
\label{sec:acoustics}

Admittedly, the study of topological insulators has drawn considerable
interest in a variety of fields, among other reasons because of the
ability of such media to feature transport that is immune to the
presence of defects~\cite{TI, TI2}. One of the most recent
venues for such studies has been in the area of mechanical and
acoustic systems, where the topological properties can inspire the
design of unconventional mechanical structures with unique elastic
and vibrational properties~\cite{Pal,Huber1,Nash,Mousavi,Wang}.
This, in turn, can lead to significant new paradigms in the
realm of energy harvesting, as well as in that of vibration
isolation~\cite{Huber2}.

One of the canonical examples that is possible to realize in this
mechanical setting is a direct analogue of a dimer in the form of
an SSH model as has been suggested e.g. in~\cite{Kane}.
This enables through its corresponding phononic band-gap
structure the emergence of a zero-frequency topological mode.
Finite (non-zero) frequency topological modes can also be
achieved~\cite{Huber3}. There have been numerous recent efforts in this direction
of harnessing topological properties of suitable mechanical
media to improve the propagation or storage
of energy. These include, among others,
the examination of edge solitons~\cite{ypma} and their ability
for nonlinear conduction in topological mechanical
insulators~\cite{vitelli,bertoldi}, the study of nonlinear
edge states that arise in  phononic lattices~\cite{ruzzene}, as well
as the examination of topological band transitions in tunable
phononic systems, under the variation of suitable (e.g. stiffness)
parameters~\cite{rajesh1,rajesh2}.


Lastly, we touch upon the theme of
topology optimization which is an important method used for many industrial and technological applications.  For structural robustness and additive manufacturing, nonlinear topology optimization \cite{topoNL1} and topology optimization for geometrically nonlinear structures \cite{topoNL2} have been recently studied.  In the former case a nonlinear elastic
model of the materials is considered along with plasticity aspects in conjunction with invoking the von Mises (yield) criterion.  In the latter case it is assumed that the structures
under consideration experience large displacement but small strain.  In particular, high-resolution topology optimized solutions are obtained for structures that are geometrically nonlinear. Invariably, the results are quite important for many engineering applications.

\section{Conclusions and Future Work}

In this chapter, we have delineated the importance of topology in a variety of physical systems and discussed the ubiquity of topological defects such as skyrmions, merons,
vortices, hopfions and monopoles in a number of distinct nonlinear
systems, e.g. chiral magnets, liquid crystals, BEC, etc.  We have also
elucidated the pervasive role of topology in nonlinear condensed
matter and photonic, as well as phononic systems.  Although we did not
discuss it here, topology in soft matter is also quite important
e.g. in topological colloids \cite{colloids}.
In addition, we illustrated the interplay of nonlinearity, topology and geometry by considering the nonlinear $\sigma$ model on simple curved manifolds.  The resulting spin configurations are sine-Gordon or double sine-Gordon solitons.

Despite the above significant progress over the last few years,
there are several important open problems related to the interplay of nonlinearity and topology.  Nonlinearity with ``fragile topology'' in a quantum system is a topic for future research.  Fragile topology (as opposed to strong topology) refers to a set of quantum phenomena that endow materials or systems with unusual properties \cite{slager2,island, bernevig}.
Examples include the misaligned layers of graphene \cite{fragile} and ``knotty'' electronic quantum states in some topological materials.  In the latter case electrons are restricted to move along certain directions. Understanding these states properly may require considerations other than K-theory \cite{ktheory}, as discussed in e.g. \cite{slager2}.

Akin to the study of soliton collisions and vortex interactions, it would be desirable to study interaction between and collision of different hopfions.  This certainly is a challenging numerical problem.

A study of one-dimensional NLD and NLW equations on (planar and space) curves and higher (two and three) dimensional such equations on curved manifolds will provide important insights into the interplay of topology, geometry and nonlinearity.  In particular, the shape and dynamics of soliton and vortex solutions will be modified by the curved geometry.

There are three fundamental (relativistic) fermions in nature: Dirac, Weyl (with zero mass) and Majorana (which are their own antiparticles and thus neutral). Similar to NLD and NLW equations there could exist a nonlinear Majorana equation; it would be intriguing to explore its soliton solutions and their dynamics. Majorana fermions also have potential applications in the growing field of (braiding-based)  topological quantum computing \cite{nayak}, especially with fault tolerance.  We note here that the nonlinear dynamics of Majorana modes has been studied using topological Josephson junctions \cite{feng}. Similarly, it was proposed that Majorana-like modes of light can also be realized in a one-dimensional array of nonlinear cavities \cite{bardyn}.

During the last few years there have been studies of a class of nonlinear models that harbor kink solitons with non-exponential tails. Collision of such kinks with non-exponential
tails, e.g. power-law \cite{ptails} or super-exponential, in 1+1 dimensional field theories is an important open question.

Recently the field of quantum time crystals \cite{wilczek} has emerged with many insights including topological considerations \cite{naka}. The role of nonlinearity in quantum time crystals with topological aspects remains an open field.


\begin{acknowledgement}

A.S. and P.G.K. acknowledge the support of the U.S.\ Department of
Energy.
Specifically, LANL is operated by Triad National Security, L.L.C.\ for
the National Nuclear Security Administration of the U.S.\ Department
of Energy under Contract No. 892333218NCA000001.
J.C.M. thanks financial support from MAT2016-79866-R project (AEI/FEDER, UE).
This material is based upon work supported by the National Science Foundation
under Grant No. DMS-1809074 (P.G.K.). Finally, P.G.K. gratefully
acknowledges
the support of the Leverhulme Trust towards a visiting fellowship
at the University of Oxford and the kind hospitality of the
Mathematical
Institute of the University of Oxford.

\end{acknowledgement}



\printindex
\end{document}